\documentclass[prd,twocolumn,floatfix,showpacs,nofootinbib,amsmath,amssymb]{revtex4}

\usepackage[latin1]{inputenc}
\usepackage[T1]{fontenc}
\usepackage{graphicx}
\usepackage[dvips]{epsfig}
\usepackage{amsmath}
\usepackage{amssymb}
\usepackage{slashed}
\usepackage{color}
\usepackage{comment}
\usepackage{hyperref}

\allowdisplaybreaks 

\begin{document}

\preprint{PRD}

\title{Infrared Behavior of Three-Point Functions in Landau Gauge Yang-Mills Theory}

\author{R. Alkofer\footnote{reinhard.alkofer@uni-graz.at}, M. Q. Huber\footnote{markus.huber@uni-graz.at}, K. Schwenzer\footnote{kai.schwenzer@uni-graz.at} \\}

\affiliation{Institut f\"ur Physik, Karl-Franzens Universit\"at Graz, Universit\"atsplatz 5, 8010 Graz, Austria}
\date{\today}

\begin{abstract}
\noindent 
Analytic solutions for the three-gluon and ghost-gluon vertices in Landau gauge Yang-Mills theory at low momenta are presented in terms of hypergeometric series . They do not only show the expected scaling behavior but also additional kinematic divergences when only one momentum goes to zero.
These singularities, which have also been proposed previously, induce a strong dependence on the kinematics in many dressing functions.
The results are generalized to two and three dimensions and a range of values for the ghost propagator's infrared exponent $\kappa$.
\end{abstract}


\maketitle


\newcommand{\mhalfo}{\frac{1}{2}}	
\newcommand{\mhalf}[1]{\frac{#1}{2}}
\newcommand{\ka}{\kappa}
\newcommand{\equ}[1]{\begin{equation} #1 \end{equation}}
\newcommand{\ba}{\begin{align}}
\newcommand{\ea}{\end{align}}	
\newcommand{\eref}[1]{Eq. (\ref{#1})}
\newcommand{\fref}[1]{Fig. \ref{#1}}
\newcommand{\ddotp}[1]{\frac{d^d #1}{(2\pi)^d}}	
\newcommand{\nnnl}{\nonumber\\}	
\newcommand{\G}[1]{\Gamma(#1)}
\newcommand{\nq}{\nu_1}	
\newcommand{\nw}{\nu_2}	
\newcommand{\nd}{\nu_3}	
\newcommand{\fig}[4]{\begin{figure}[#1]\centering\epsfig{file=#3}\caption{#4}\label{#2}\end{figure}}

\section{Introduction}\label{sec:intro}

During the last ten years the infrared (IR) behavior of Yang-Mills Green functions was studied intensively. 
Especially the propagators are of interest as they are linked to the confinement problem via the scenarios of Kugo-Ojima \cite{Kugo:1979gm} and 
Gribov-Zwanziger \cite{Gribov:1977wm,Zwanziger:1991gz}. 
Both predict in Landau gauge an IR enhanced ghost propagator that is responsible for the appearance of long-range forces, whereas the gluon propagator should vanish 
(or be at least finite for the former). Indeed such behavior was found by functional methods 
(Dyson-Schwinger equations (DSEs) \cite{Smekal:1997is,vonSmekal:1997is,Lerche:2002ep,Zwanziger:2001kw,Alkofer:2000wg,Fischer:2008uz} as well as 
Renormalization Group (RG) equations \cite{Pawlowski:2003hq,Fischer:2008uz}).

The DSEs of Yang-Mills theory in Landau gauge allow two different kind of solutions, called the scaling and the decoupling solution in ref.\ \cite{Fischer:2008uz}. 
The latter has been addressed in refs.\ \cite{Boucaud:2008ji,Aguilar:2008xm,Dudal:2008sp,Fischer:2008uz,Alkofer:2008jy,Schwenzer:2008vt}. 
Especially, for this solution the behavior of vertices was determined in \cite{Alkofer:2008jy} as trivial, i.e.\ the dressing functions are at most constant in the IR, 
and there is no IR enhancement. Furthermore, the decoupling solution does break global BRST and global gauge invariance in the confining phase \cite{Fischer:2008uz}, 
whereas the scaling solution can provide a picture for confinement in Landau gauge \cite{Alkofer:2006gz,Alkofer:2008tt}. 
Therefore we will concentrate on the scaling solution, as described in sec. II, in this paper.

Besides the propagators also the vertices play an important role in Landau gauge Yang-Mills theory. Whereas the ghost-gluon vertex is IR regular the three-gluon vertex 
is strongly IR divergent \cite{Alkofer:2004it}. In fact, this happens for all dressed gluonic vertices in the uniform limit that all momenta vanish simultaneously. 
Thereby the IR behavior of any diagram in a DSE is characterized by its bare vertex  \cite{Huber:2007kc}. 
A recent power counting analysis showed that the three-gluon vertex can feature additional kinematic singularities 
when only a single gluon momentum vanishes \cite{Alkofer:2008jy}. As has been explained in detail in ref.\ \cite{Alkofer:2008tt}, the importance of the vertex functions is even more 
pronounced in the quark sector: Here the dressed quark-gluon vertex may provide an explicit mechanism for quark confinement in Landau gauge QCD. In this vertex
it is a strong soft-gluon divergence in quenched QCD that is able to generate the necessary interaction strength to permanently confine quarks.
This strong divergence is in turn self-consistently induced by a contribution involving effectively the strongly IR divergent three-gluon vertex.

Here we present a detailed study of the IR behavior of the three-point vertices (the three-gluon and the ghost-gluon vertex) in Landau gauge Yang-Mills theory. 
Within a semi-perturbative approximation we provide explicit analytic results for the full kinematic dependence in the IR limit that confirm and exemplify 
the results of our previous power counting analysis \cite{Alkofer:2008jy}. These results allow to develop better ans\"atze for these vertices in subsequent investigations.
This will provide more reliable results for the gluon propagator and the full momentum dependence of Yang-Mills vertices 
and in particular for the analysis of the quark-gluon vertex.

We first review the IR fixed point structure of Landau gauge Yang-Mills theory obtained from a pure power counting analysis in sec.\ \ref{sec:IR-beh-vert}. 
To go beyond this we present an analytic approximation scheme that covers the leading IR contributions to the three-point vertices in sec. \ref{se:3P-Int}, 
where we also present general results for the triangle graphs with dressed propagators in the IR limit and show how they can be evaluated in the Euclidean case. 
The explicit results for the individual tensor structures of the three-gluon vertex and the ghost-gluon vertex are presented in secs.\ 
\ref{sec:results} and \ref{sec:resultsGhGV}, respectively. Technical aspects like the tensor decomposition are deferred to two appendices.

\section{Infrared Behavior of Yang-Mills Green Functions}\label{sec:IR-beh-vert}

Neglecting the trivial color factor $\delta_{ab}$, the gluon and ghost propagators are parametrized as
\begin{equation}
D_{\mu\nu}(p^2)=\left( \delta_{\mu \nu}-\frac{p_\mu p_\nu}{p^2} \right) 
\frac{Z(p^2)}{p^2} \quad
\end{equation}
and
\begin{equation}
\quad D_G(p^2)=-\frac{G(p^2)}{p^2}.
\end{equation}
For very small momenta it is expected that the dressing functions behave power-like:
\begin{equation}\label{eq:power-laws-props}
Z^{IR}(p^2)=c_{0,2} \cdot(p^2)^{2\ka}, \quad \quad G^{IR}(p^2)= c_{2,0} \cdot(p^2)^{-\ka}.
\end{equation}
The exponents of the squared momentum are called infrared exponents and depend only on one parameter, $\ka$. 
An approximate numerical value $\kappa \approx 0.5953$ can be determined from the IR consistency relation between the ghost and gluon DSEs \cite{Lerche:2002ep,Zwanziger:2001kw}. 
Recently a similar value $\kappa\approx0.57$ was obtained in a strong coupling lattice analysis \cite{Sternbeck:2008mv}. 
In agreement with the Kugo-Ojima and the Gribov-Zwanziger scenarios the ghost propagator is divergent in the infrared, whereas the gluon propagator is slightly vanishing
like $(p^2)^{0.2}$. 
(NB: For an analysis of the $\ka$-dependence of the propagators see refs.\ \cite{Lerche:2002ep,Huber:2007kc}.) 
The three-gluon vertex dependence was investigated in refs.\ \cite{Alkofer:2004it,Huber:2007kc}. 
In the uniform limit, where all momenta approach zero, also the dressing functions of vertices are given by general power laws. 
Each vertex with $2n$ ghosts and $m$ gluons then has at least one dominant tensor that behaves as 
$c_{2n,m}(\ka,\lbrace p_i^2/p^2\rbrace) \cdot (p^2)^{\delta_{2n,m}}$ with some prefactor $c_{2n,m}$, provided no non-trivial cancelations occur. 
$\delta_{2n,m}$ is hereby the uniform IR exponent that determines the IR behavior in terms of one common momentum scale $p^2$.

The general expression for the (uniform) IR exponent in $d$ dimensions is \cite{Alkofer:2004it,Huber:2007kc,Alkofer:2007hc}
\equ{\label{eq:IR-exp}
\delta_{2n,m}=(n-m)\ka+(1-n)\left(\mhalf{d}-2\right).}
One possible way to obtain \eref{eq:IR-exp} is via the non-renormalization of the ghost-gluon vertex in Landau gauge \cite{Taylor:1971ff}, which is connected to a finite IR dressing of the vertex: $\delta_{2,1}=0$. 
This has been confirmed with DSE methods for $d=3,4$ \cite{Schleifenbaum:2004id} and calculations on the lattice for $d=2,3,4$ \cite{Cucchieri:2004sq,Cucchieri:2006tf,Ilgenfritz:2006he,Maas:2007uv,Cucchieri:2008qm}.
For higher vertex functions, e.g.\ the 1PI five-gluon vertex, a skeleton expansion was used in refs.\ \cite{Alkofer:2004it,Huber:2007kc,Alkofer:2007hc}. 
A feature of this expansion is that all orders possess the same IR exponent \cite{Alkofer:2004it}, which is true also for two and three dimensions \cite{Huber:2007kc,Alkofer:2007hc}. 
As a matter of fact the IR exponent is equal for all diagrams with the same type of bare vertex \cite{Huber:2007kc}. 
For the three-gluon vertex DSE diagrams containing a bare ghost-gluon vertex turn out to have an IR exponent of $-3\ka$. 
Those with a bare three- or four-gluon vertex have $0$ and $\ka$, respectively. Another feature of this solution is the IR fixed-point of the running coupling \cite{vonSmekal:1997is,Alkofer:2004it,Fischer:2002hn,Alkofer:2002ne,Huber:2007kc}.

In all these considerations it was assumed that vertex functions feature only singular behavior when all momenta vanish uniformly. Taking into account that vertex functions depend on several independent momenta, a more refined picture emerges that does not alter but extend the previously described one \cite{Alkofer:2008jy}. The dependence on different momenta is encoded in the explicit dependence of the vertex dressing functions on all independent momenta. A general vertex function can then show power law scaling with different IR exponents $\delta_{i,t}$ depending on the momentum configuration $p_i^2\left(q_{1}^{2},\cdots,q_{n}^{2}\right)$ that vanishes:
\begin{align}
&\Gamma^{\mu_{1}\cdots\mu_{m}}\left(q_{1},\cdots,q_{n}\right)=  \nnnl
&\quad=\sum_{t}\sum_{i}c_{i,t}\left(q_{1}^{2}/p_{i}^{2},\cdots,q_{n}^{2}/p_{i}^{2}\right)\times \nnnl
&\quad\times\left(p_{i}^{2}\left(q_{1}^{2},\cdots,q_{n}^{2}\right)\right)^{\delta_{i,t}}T_{t}^{\mu_{1}\cdots\mu_{m}}\left(q_{1},\cdots,q_{n}\right). \nonumber \label{eq:general-scaling}\end{align}
Such a non-trivial momentum dependence is also found in the analytic IR solution and is in perfect agreement with the power counting analysis performed in ref.\ 
\cite{Alkofer:2008jy},
where all IR exponents for the primitively divergent Green functions were determined. For the three-point functions three new IR exponents for the case when one momentum 
becomes small compared to the other two were introduced.
In the following we present explicit calculations that confirm the related previously obtained results of ref.\ \cite{Alkofer:2008jy}.

\begin{table}[h]
\begin{center}
\begin{tabular}{|c|c||c|c||c|c|c|}
\hline 
$\delta_{gh}$ & $\delta_{gl}$ & $\delta_{gg}^{u}$ & $\delta_{3g}^{u}$ & $\delta_{gg}^{gh}$ & $\delta_{gg}^{gl}$ & $\delta_{3g}^{gl}$\tabularnewline
\hline 
$-\kappa$ & $2\kappa$ & $0$ & $-3\kappa$ & $0$ & $\min(\tfrac{3}{2}-2\kappa,0)$ & $\min(1-2\kappa,0)$\tabularnewline
\hline
\end{tabular}
\end{center}
\caption{\label{tab:IR-exps}The IR exponents for the ghost propagator, the gluon propagator, the ghost-gluon and the three-gluon vertices with all momenta going to zero, the ghost-gluon vertex with a ghost momentum going to zero, the ghost-gluon and the three-gluon vertices with a gluon momentum going to zero. If not denoted otherwise, we use $\ka=0.5953\ldots$ in this article.}
\end{table}

\begin{widetext}
\begin{figure*}
\begin{center}
\includegraphics[width=\textwidth]{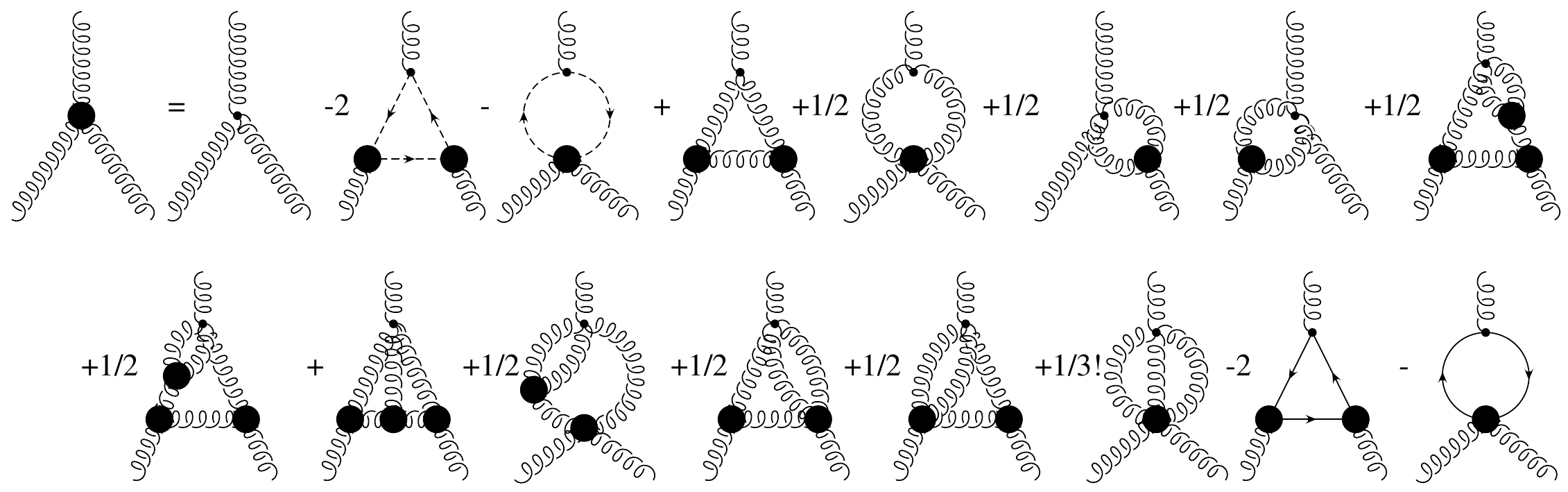}
\epsfig{file=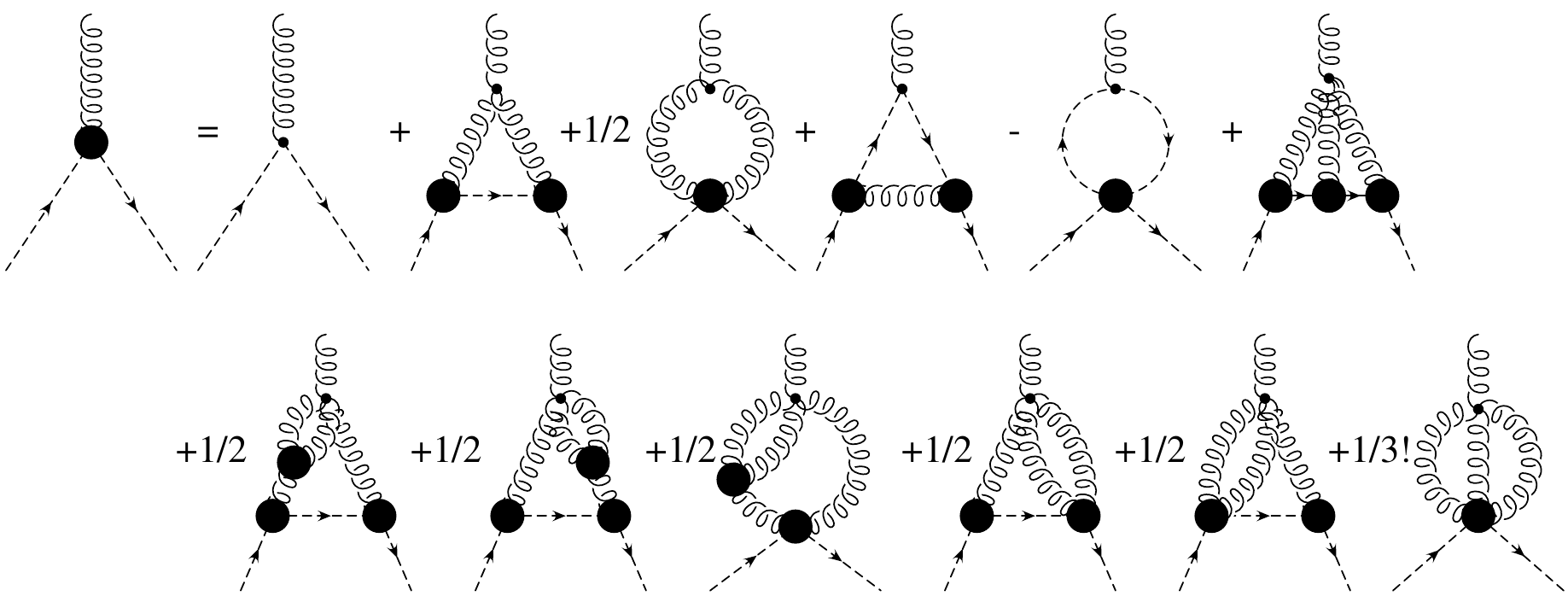,width=\textwidth}
\caption{\label{fig:3p-DSEs}The DSEs for the three-gluon and ghost-gluon vertices. They were derived using the algorithm explained in ref. \cite{Alkofer:2008nt}.}
\end{center}
\end{figure*}
\end{widetext}

The IR analysis of the three-gluon vertex DSE \cite{Alkofer:2004it} identifies the diagrams containing ghosts as the IR leading ones. The second ghost diagram of the 
three-gluon vertex DSE in \fref{fig:3p-DSEs} can be expanded using a skeleton expansion. Here the lowest diagram possesses two loops. The first order of the skeleton 
expansion is therefore the left ghost diagram in \fref{fig:3p-DSEs}, called ghost triangle in the following. The same analysis for the ghost-gluon vertex DSE identifies the fourth diagram 
on the right hand side of \fref{fig:3p-DSEs}, called ghost-ghost-gluon triangle, as the leading first order diagram. Both these diagrams are calculated in the subsequent sections. The complete DSEs as given in \fref{fig:3p-DSEs} can be derived by the algorithm explained in \cite{Alkofer:2008nt}, which is implemented in the \textit{Mathematica} package \textit{DoDSE}.

Two notes on the validity of the results of the presented calculations are in order: Because we do not solve a full system of DSEs but rather calculate single diagrams for low external momenta we only can provide qualitative solutions.
Furthermore, we treat only the Yang-Mills sector. This is sufficient for an analysis of QCD in the quenched approximation \cite{Alkofer:2008tt}
as well as in the unquenched case \cite{Schwenzer:2008vt}.

\section{The Generalized Triangle Integrals}\label{se:3P-Int}

\begin{figure*}
\centering
\epsfig{file=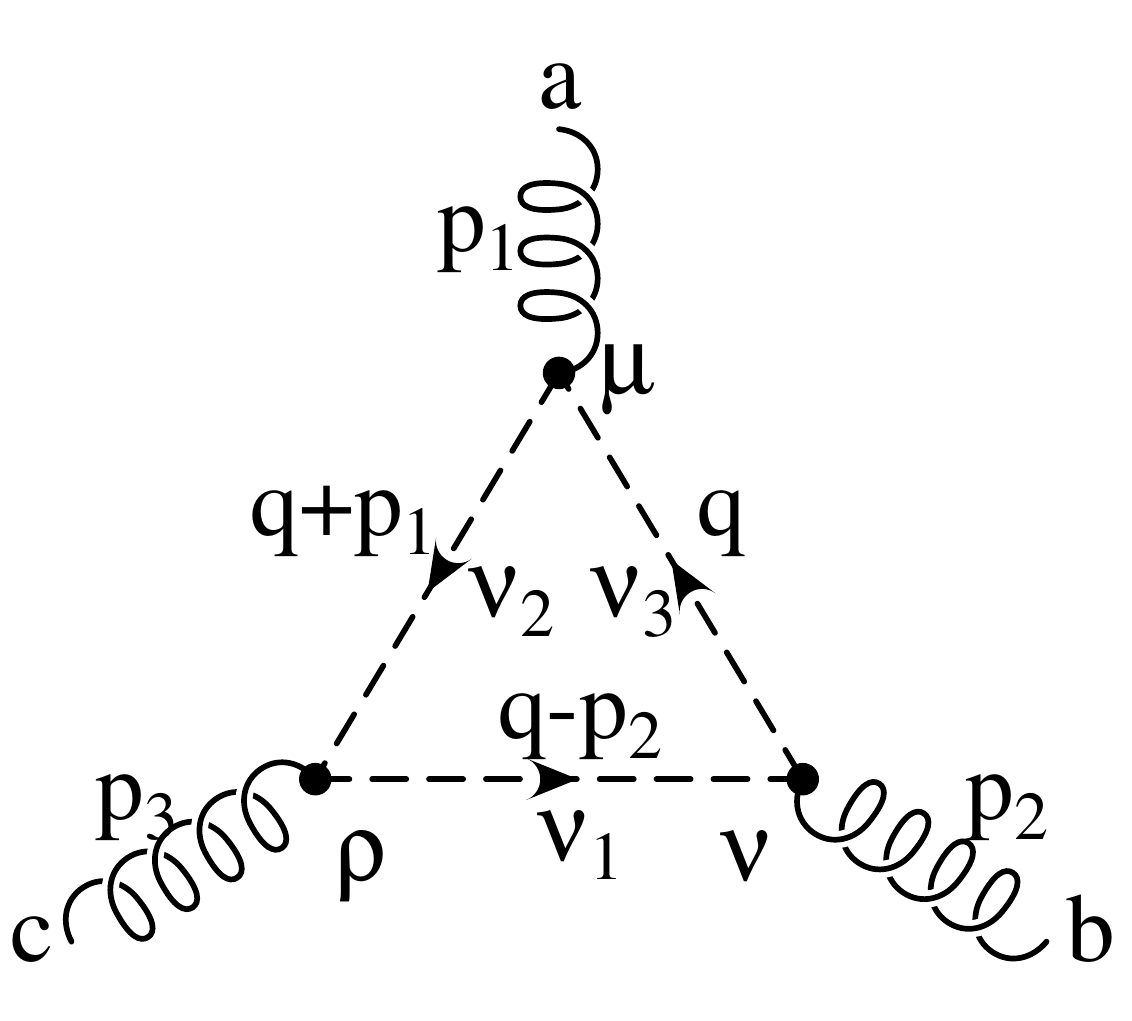,width=6cm}
\epsfig{file=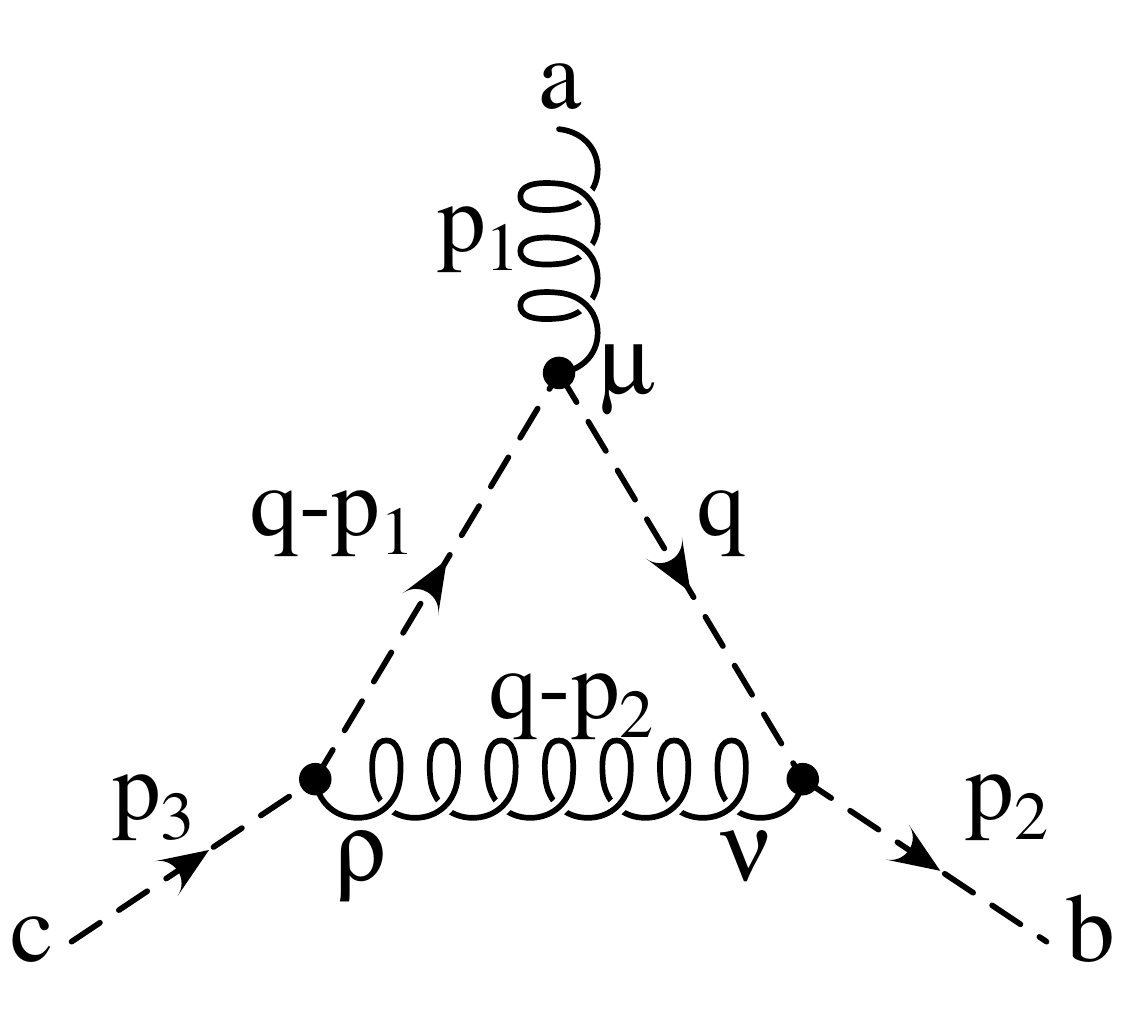,width=6cm}
\caption{Momentum routing for the ghost and the ghost-ghost-gluon triangles.}
\label{fig:triangle}
\end{figure*}

We will calculate the two diagrams given in \fref{fig:triangle}. In the following it is understood that the DSEs have been renormalized by a suitable subtraction scheme \cite{vonSmekal:1997is} and no terms divergent in the ultraviolet are present. Thus all quantities are to be considered ultraviolet finite even without writing down explicitly the renormalization constants. Our presentation will mainly focus on the ghost triangle as the procedure is quite similar for the ghost-gluon vertex.
Using bare ghost-gluon vertices, $i\,g\,f^{abc}\,q_{\mu}$ with $q_{\mu}$ being the outgoing ghost momentum, but dressed propagators, the ghost triangle correction in the renormalized three-gluon DSE yields
\begin{align}
\Gamma&_{\mu \nu \rho}^{\Delta}(p_1,p_2;\ka;d)= i\,f^{abc}\mhalf{N_c\,c_{2,0}^3\,g^3} \times \nnnl
	& \quad \times \int \ddotp{q} (q+p_1)_\mu(q-p_2)_\rho q_\nu \times \nnnl
	& \quad \times  
		\frac{G((q+p_1)^2)}{(q+p_1)^2} \frac{G((q-p_2)^2)}{(q-p_2)^2}\frac{G(q^2)}{q^2}.
\end{align}
For small external momenta $p_1$ and $p_2$ the integral will be dominated by the singularities of the IR ghost propagators  (\eref{eq:power-laws-props}). It can be decomposed in one vector and two tensor integrals of rank two and one of rank three,

\begin{widetext}
\begin{align}
	^{IR}\Gamma&_{\mu \nu \rho}^{\Delta}(p_1,p_2;\ka;d)= i\,f^{abc}\mhalf{N_c\,c_{2,0}^3\,g^3} 
\int \ddotp{q} (q+p_1)_\mu(q-p_2)_\rho q_\nu 
		((q+p_1)^2)^{-\ka-1}((q-p_2)^2)^{-\ka-1}(q^2)^{-\ka-1}=\nnnl
	& = i\,f^{abc}\mhalf{N_c\,c_{2,0}^3\,g^3} \big \lbrace- p_{1_{\mu}} p_{2_{\rho}} I_{\nu}(p_1,p_2,p_3;1+\ka,1+\ka,1+\ka;d) + \nnnl
	& \quad + p_{1_{\mu}}I_{\nu \rho}(p_1,p_2,p_3;1+\ka,1+\ka,1+\ka;d) + I_{\mu \nu \rho}(p_1,p_2,p_3;1+\ka,1+\ka,1+\ka;d) -\nnnl
	& \quad -p_{2_{\rho}} I_{\mu \nu}(p_1,p_2,p_3;1+\ka,1+\ka,1+\ka;d))\big \rbrace,
\end{align}
where the tensor integrals are defined as
\begin{align}
I_{\mu_1 \mu_2 \ldots}(p_1,p_2,p_3;\nq,\nw,\nd;d):=\int \frac{d^d q}{(2\pi)^d} \frac{q_{\mu_1}q_{\mu_2}\ldots}{((q \pm p_2)^2)^{\nu_1} ((q \pm p_1)^2)^{\nu_2} (q^2)^{\nu_3}}.
\end{align}
The signs in the denominators depend on the momentum routing, but the solution for the scalar integrals is independent of them.
The positive factor $c_{2,0}$ in the ghost dressing function and the coupling $g$ are to be considered as unknown constants because we have no scale to determine physical values for them. Therefore the integrals can only give information about the qualitative behavior of the three-gluon vertex as mentioned above. The factor $-N_c/2$ stems from the color algebra. In all calculations only the term in braces was considered.
For the ghost-ghost-gluon triangle the same procedure was performed.

Using a method due to Davydychev \cite{Davydychev:1991va} we can decompose the tensor integrals into scalar ones,
\begin{align}\label{eq:scalar-3Point-Integral}
I(p_1,p_2,p_3;\nq,\nw,\nd;d):=\int \frac{d^d q}{(2\pi)^d} \frac{1}{((q \pm p_2)^2)^{\nu_1} ((q \pm p_1)^2)^{\nu_2} (q^2)^{\nu_3}}.
\end{align}
\end{widetext}
The advantage of this method, described in more detail in App. \ref{sec:tensorBases}, is that it involves only a minimal number of scalar integrals in higher dimensions. In contrast, a direct decomposition into four-dimensional integrals involves a vast number of different scalar integrals that mostly cancel mutually and yields equivalent results as we have checked in certain cases.

The remaining scalar integrals involve denominators with non-integer exponents. 
The solution for such integrals was obtained both via a Mellin-Barnes representation for occurring Gaussian hypergeometric functions \cite{Boos:1987bg} and within the Negative Dimensions Integration Method (NDIM) \cite{Anastasiou:1999ui}. The result in general dimension, involving Appell's hypergeometric series $F_4$, is given by
\begin{widetext}
\begin{align}\label{eq:I-final}
&I(p_1,p_2,p_3;\nq,\nw,\nd;d)=\frac{1}{(4\pi)^\mhalf{d}}(p_3^2)^{\mhalf{d}-\nw-\nd-\nq}\times\nnnl
  &\quad \times \Bigg\lbrace \left(\frac{p_1^2}{p_3^2}\right)^{\mhalf{d}-\nw-\nq}\left(\frac{p_2^2}{p_3^2}\right)^{\mhalf{d}-\nw-\nd} \frac{\G{\mhalf{d}-\nw}\G{-\mhalf{d}+\nw+\nq}\G{-\mhalf{d}+\nw+\nd}}
      {\G{\nw}\G{\nd}\G{\nq}}\times\nnnl
      &\qquad \times F_4\left(d-\nw-\nd-\nq,\mhalf{d}-\nw;1+\mhalf{d}-\nw-\nq,1+\mhalf{d}-\nw-\nd;\frac{p_1^2}{p_3^2},\frac{p_2^2}{p_3^2}\right)+\nnnl
  &\quad+\frac{\G{\mhalf{d}-\nw-\nq}\G{\mhalf{d}-\nw-\nd}\G{-\mhalf{d}+\nw+\nd+\nq}}
      {\G{\nd}\G{\nq}\G{d-\nw-\nd-\nq}}\times\nnnl
      &\qquad \times F_4\left(\nw,-\mhalf{d}+\nw+\nd+\nq;1-\mhalf{d}+\nw+\nq,1-\mhalf{d}+\nw+\nd;\frac{p_1^2}{p_3^2},\frac{p_2^2}{p_3^2}\right)+\nnnl
  &\quad+\left(\frac{p_1^2}{p_3^2}\right)^{\mhalf{d}-\nw-\nq}
       \frac{\G{\mhalf{d}-\nq}\G{\mhalf{d}-\nw-\nd}\G{-\mhalf{d}+\nw+\nq}}
      {\G{\nw}\G{\nq}\G{d-\nw-\nd-\nq}}\times\nnnl
      &\qquad \times F_4\left(\mhalf{d}+\nq,-\nd;1+\mhalf{d}+\nw+\nq,1-\mhalf{d}-\nw-\nd;\frac{p_1^2}{p_3^2},\frac{p_2^2}{p_3^2}\right)+\nnnl
  &\quad+\left(\frac{p_2^2}{p_3^2}\right)^{\mhalf{d}-\nw-\nd} \frac{\G{\mhalf{d}-\nd}\G{\mhalf{d}-\nw-\nq}\G{-\mhalf{d}+\nw+\nd}}
      {\G{\nw}\G{\nd}\G{d-\nw-\nd-\nq}}\times\nnnl
      &\qquad \times F_4\left(\nq,\mhalf{d}-\nd;1-\mhalf{d}+\nw+\nq,1+\mhalf{d}-\nw-\nd;\frac{p_1^2}{p_3^2},\frac{p_2^2}{p_3^2}\right)
\Bigg\rbrace
\end{align}
\end{widetext}
with Appells' series $F_4$ defined as
\begin{equation}
\label{eq:F4-def}
F_4(a,b;c,d;x,y)=\sum_{m,n=0}^{\infty}\frac{(a,m+n)(b,m+n)}{(c,m)(d,n)}\frac{x^m}{m!}\frac{y^n}{n!}.
\end{equation}
The so-called Pochhammer symbol $(a,n)$ is given in \eref{eq:def-PochhammerWithGamma}.
When interested in the Euclidean momentum region this result is not directly applicable, since the defining series does not converge there. Its region of convergence, given by
\begin{align}
\label{eq:F4-conv}
&\sqrt{|x|}+\sqrt{|y|}<1,
\end{align}
is shown in \fref{fig:roc} as the dark gray region. 
Via known expressions \cite{Erdelyi:1953ht,Gradshteyn:1980is,Exton:1994de} this series can be continued analytically to the light gray areas which correspond to solutions for the two other possible ways of forming momentum ratios in eq. (\ref{eq:I-final}).

In contrast, using momentum and energy conservation for the three external momenta $p_1$, $p_2$ and $p_3$, the Euclidean region is defined by ($\alpha$ is the angle between $p_2$ and $p_3$)
\equ{
p_1^2=(-p_2-p_3)^2
=p_2^2+p_3^2+2|p_2||p_3|\cos(\alpha)
}
leading for $x=p_1^2/p_3^2$ and $y=p_2^2/p_3^2$ to the expression
\begin{align}\label{eq:roc-Eucl}
x+y-2\sqrt{x\,y}&<1<x+y+2\sqrt{x\,y}.
\end{align}
This region is precisely the white part in \fref{fig:roc}. A series representation of Appell's function $F_4$ convergent in a part of the Euclidean region is given in Appendix \ref{sec:anal-cont-F4}. Employing symmetries and other, simpler continuation formulae one can cover the whole Euclidean region.
\fig{}{fig:roc}{roc,width=0.7\linewidth}{The regions of convergence for different series representations of Appell's function $F_4$.}

Recently it was claimed \cite{Suzuki:2007da, Suzuki:2008hm} that \eref{eq:I-final} is not correct and the number of Appell's series should be reduced to three. However, the procedure performed to this end only amounts to a standard analytic continuation of two Appell's series to get only one. This is not per se forbidden, but the new Appell's series has a region of convergence (the upper light gray region in \fref{fig:roc}) that does not intersect that of the two remaining ones (the dark gray region), so that the total result does not converge. We therefore do not see, how such a reduction is of any help.

An important feature of the three-point integral solution are divergences at some kinematic points. Appell's function $F_4$ may possess poles at $(\infty,\infty)$, $(0,1)$ and $(1,0)$, but it can also be finite or vanishing depending on the parameters. Another reason for kinematic divergences are the prefactors in \eref{eq:I-final}, which diverge for certain values of the parameters in case of collinear momenta, i.e. one momentum is zero. Approaching these poles is equivalent to letting only one momentum vanish and is described by the non-uniform IR exponents when only a single momentum vanishes. We call these points asymmetric points. Since the final values of the IR exponents are a result of the intricate interplay between the divergences in the prefactors and of the Appell's functions, we extracted all values numerically and checked some of them analytically.
As an example the quantitative result for the scalar integral $I(p_1^2,p_2^2,p_3^2=1;1+\ka,1+\ka,1+\ka;4)$ is shown in \fref{fig:scal-int}.
The analytic result is compared to a result from a numerical integration and shows good agreement. The general features of this function are that it rises monotonically from the symmetric point $(1,1)$ towards the boundary of the Euclidean region. At the boundary it has singularities at $(1,0)$, $(0,1)$ and $(\infty ,\infty)$ and saddlepoints in between.
One can clearly see these divergences when one momentum becomes small.


A remark on the asymmetric point: Under certain circumstances one can evaluate the integral exactly at the point $p_i=0$ by use of the two-point integral,
\begin{align}\label{eq:I2}
&\int d^dq\frac{1}{(q^2)^{\mu_1}}\frac{1}{((q\pm p)^2)^{\mu_2}}=\nnnl
&=(p^2)^{\frac{d}{2}-\mu_1-\mu_2}\pi^{\frac{d}{2}}
	\frac{\Gamma(\frac{d}{2}-\mu_1)\Gamma(\frac{d}{2}-\mu_2)\Gamma(\mu_1+\mu_2-\frac{d}{2})}
		{\Gamma(\mu_1)\Gamma(\mu_2)\Gamma(d-\mu_1-\mu_2)},
\end{align}
where the conditions
\begin{align}\label{eq:I2pCond}
d/2-\mu_1-\mu_2 \leq0,\quad d/2-\mu_1\geq0,\quad d/2-\mu_2\geq0
\end{align}
have to be fulfilled for IR and UV convergence.
One can directly show this expression to be equivalent to the three-point solution at the asymmetric point for $1-\nq-\nd>0$. Most easily this can be seen for the case $p_2^2\rightarrow 0$ by setting it to zero in \eref{eq:I-final}. Thereby the Appell's series become one-dimensional Gaussian hypergeometric series $_2F_1$ and two terms vanish if the exponents of $p_2^2$ are greater than zero ($d/2-\nq-\nd>0$). For $p_1^2=p_3^2$ the Gaussian series have unit argument and converge only for $1-\nq-\nd>0$. This means that for $d>2$ the latter condition is stronger. If it is fulfilled the two-point and three-point integrals coincide, which can be seen by setting $\mu_1=\nu_1+\nu_3$ and $\mu_2=\nu_2$. The second condition in \eref{eq:I2pCond} corresponds then to $d/2-\nu_1-\nu_3\geq0$. The case when this condition is not true yields the kinematic singularities treated in this article and the two-point integral cannot be used for calculating this limit.

\fig{}{fig:scal-int}{scalar-plot,width=0.9\linewidth}{The scalar integral \eref{eq:I-final} in the Euclidean momentum region at fixed momentum $p_3^2$. Overlayed are both the analytic solution (black contour lines) and the result of a numerical integration (gray contour lines) for comparison. The slight deviations near the singularities result from the slow convergence of the numerical integration routine.}

As explained in Sec. \ref{sec:IR-beh-vert}, the power counting analysis only gives the IR exponent for the most divergent dressing function that is possible. Neither do specific results have to match the exponents nor can less divergent, finite or even vanishing dressing functions be excluded. Indeed the results presented below show that the combination of Appell's functions in the scalar integrals can lead to the realization of these additional possibilities.

We would like to mention that in principle the investigation of the four-gluon vertex is possible in an analogous way. But although NDIM enables one to derive a solution for the ghost-box integral, the situation is more involved: Without approximations or constraints on the external momenta fivefold hypergeometric sums emerge, which are technically much more difficult to handle than a double sum. Furthermore we expect that an analytic continuation will be necessary as well, but we know of none at the moment. Apart from these problems with the integrals the tensor structure is more complicated since the four-gluon vertex is constructed of a tensor of rank four in Lorentz space. Taking also into account that a solution in one region of convergence consists of more than four series\footnote{The complete (naive) solution for the three-point integral as derived with NDIM comprises twelve series (three regions with four series each), whereas for the four-point integral this number is 162.} it is evident that in every step a multiple of the amount of work is necessary than for the three-point integral. Nevertheless calculations have been done in refs. \cite{Kellermann:2007dt,Kellermann:2008iw} for a given momentum configuration.

\section{Infrared Solution of the Three-Gluon Vertex}\label{sec:results}

A perturbative solution for the three gluon vertex in arbitrary linear covariant gauge and dimension valid in the UV regime is given in \cite{Davydychev:1996pb}. 
Here we give a semi-perturbative result that involves fully dressed propagators valid in the IR regime. With a general solution to the massless three-point integral 
(\eref{eq:I-final} together with the results from the analytic continuation presented in Appendix \ref{sec:anal-cont-F4}) and Davydychev's method \cite{Davydychev:1991va} 
for calculating tensor integrals (see Appendix \ref{sec:tensorBases}) at hand we are able to present here the result of the IR dominant part of the three-gluon and 
ghost-gluon vertices in the complete Euclidean momentum regime in the minimal tensor basis provided by the employed computation method \cite{Davydychev:1991va}.
Furthermore, we can give the overlap of the vertex with arbitrary tensors and do so for the transverse part of the vertex for a comparison with lattice data.

The results are obtained completely analytically and allow in principle to get the IR exponents directly. However, due to the involved form of the integrals it is much faster to extract them by numerical fits. We did this for $0< \ka < 1$ and $d=2,3,4$ dimensions and present the results in table \ref{tab:IR-Exps-Num-kappa}, which also shows that the dependence on $\ka$ is more involved as originally anticipated, since besides the additional term arising in two and three dimensions it is also possible that the dressing functions become constant. In the three-dimensional plots and Fig.~\ref{fig:D1001Small} we used $\ka=0.5953\ldots$, which is the best currently available value \cite{Zwanziger:2001kw,Lerche:2002ep}.

 

\subsection{General Kinematic Dependence}

\begin{figure}
\begin{center}
\epsfig{file=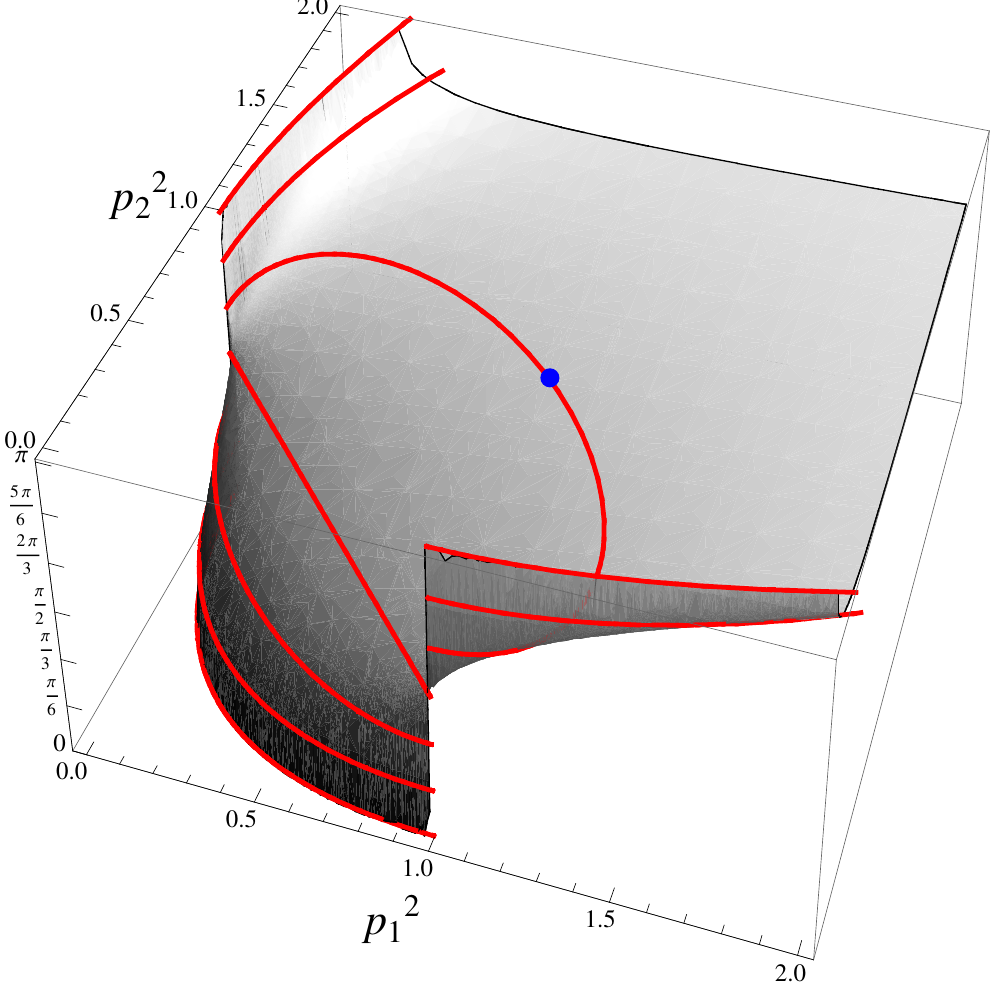,width=0.8\linewidth}
\end{center}
\caption{\label{fig:angle}For $p_3^2=1$ the angle between $p_1$ and $p_2$, which is given in general by $\cos{\alpha}=(1-x-y)/(2\sqrt{x\,y})$. The boundaries represent the collinear case $p_1\!\sim\!p_2$ and the dot lying on the $2\pi/3$ contour line represents the symmetric configuration $p_1^2\!=\!p_2^2\!=\!p_3^2$.
}

\end{figure}

As explained in Appendix \ref{sec:tensorBases}, since we only take into account the tree-level tensor of the ghost-gluon vertex, for the IR part of the three-gluon vertex only ten of the possible 14 tensors arise:
\begin{align}
\Gamma_{\mu \nu \rho}^{\Delta}(p_1,p_2,p_3)=\sum_{i=1}^{10} E_i(p_1,p_2,p_3;\ka;d) \tau^i_{\mu \nu \rho}(p_1,p_2,p_3)
\end{align}
with dressing functions $E_i$ and tensors $\tau^i_{\mu \nu \rho}$ as given in table \ref{tab:IR-Exps-Num}.
The solution involving the scalar integrals as given in eq. (\ref{eq:I-final}) consists only of hypergeometric functions which can be computed, if converging series representations can be found. Appendix \ref{sec:anal-cont-F4} provides such representations for the three-gluon vertex. The calculation of the series was performed with \textit{Mathematica 6} \cite{Wolfram:1999}. Convergence of the series is very good except at the boundaries of the Euclidean region. The free variables of the solution are $p_1^2$, $p_2^2$ and $p_3^2$, which are combined to $x=p_1^2/p_3^2$ and $y=p_2^2/p_3^3$. Due to the symmetry the full solution is contained in the unit square and is then parametrized by one unbounded scale and two bounded momentum ratios. To each point in the $x$-$y$-plane corresponds an angle between $p_1$ and $p_2$ which is depicted in \fref{fig:angle} to show the connection to another parametrization of the independent quantities employing two absolute values of momenta and the enclosed angle.
A change of $p_3^2$ corresponds to moving the kinematic point in the $(x,y)$-plane along radial lines, whereas scaling all momenta by the same factor does not change it. Thus uniform scaling $p_1^2 \sim p_2^2 \sim p_3^2 \equiv p^2$ with all ratios fixed and finite does not depend on the hypergeometric functions, where only $x$ and $y$ appear, but only on the prefactor in \eref{eq:I-final}. Taking into account the momenta from the bare ghost-gluon vertices and the anomalous scaling of the ghost propagators this gives in four dimensions
\begin{equation}
(p^2)^{\tfrac{d}{2}-\nu_1-\nu_2-\nu_3} \to (p^2)^{2-3(1+\kappa)+3\tfrac{1}{2}}=(p^2)^{\tfrac{1}{2}-3\kappa}
\end{equation}
For the three-dimensional plots we fixed $p_3^2=1$. This sets $x$ and $y$ equal to $p_1^2$ and $p_2^2$. In the following we will treat the four-dimensional case with $\ka=0.5953$, except where denoted otherwise.

Although the full solution of the three-gluon vertex is fully symmetric the choice of a tensor basis can hide this symmetry. The reason for choosing the solution of the integrals with $p_3^2$ in the denominator of the variables of the $F_4$ function is that our basis depends only on $p_1$ and $p_2$. Having the same independent momenta for the basis and the variables of the Appell's function explicitly reveals some symmetries under exchange of the two momenta in the plots, e. g. the first and the second scalars corresponding to the tensors $p_{1_{\mu}}p_{1_{\nu}}p_{1_{\rho}}/p_1^2$ and $p_{2_{\mu}}p_{2_{\nu}}p_{2_{\rho}}/p_2^2$. The sign is different because of the color factor.

\begin{figure*}
\begin{minipage}[c]{0.45\linewidth}
\epsfig{file=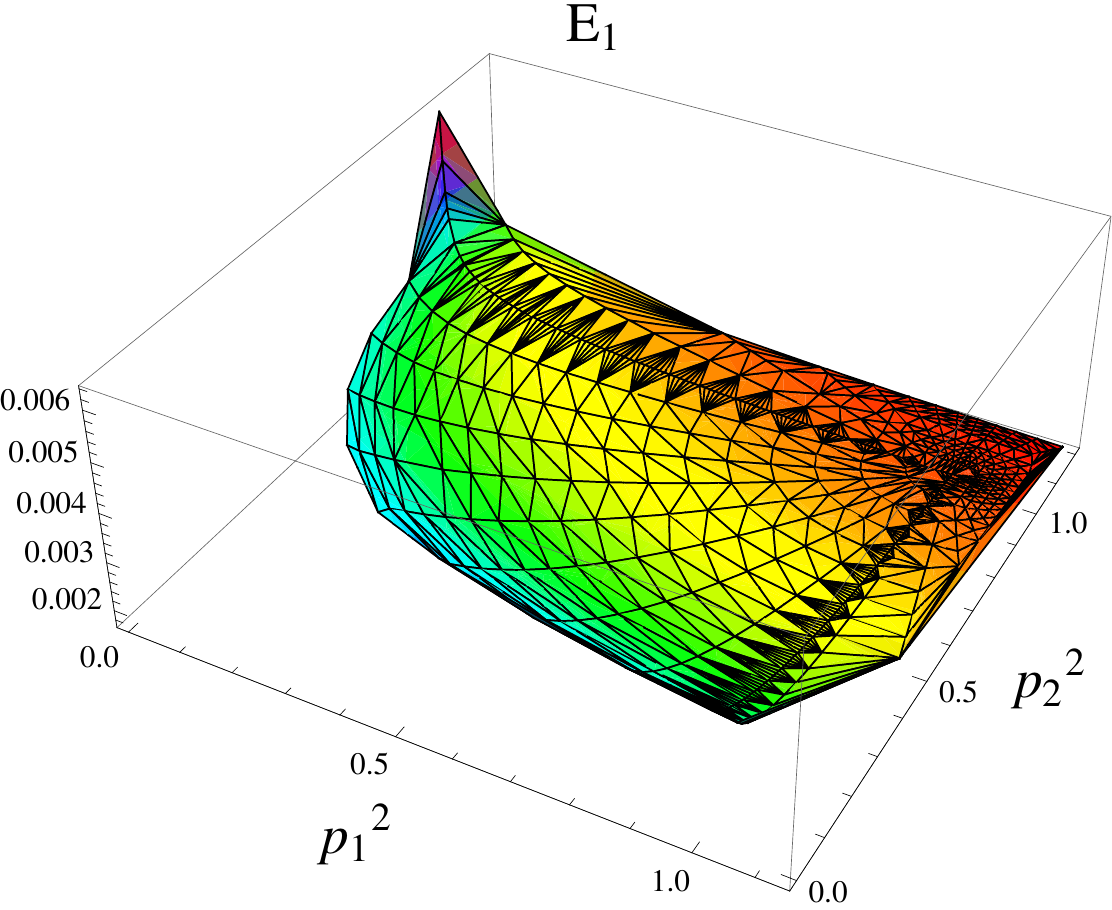,width=\linewidth}
\end{minipage}
\begin{minipage}[c]{0.45\linewidth}
\epsfig{file=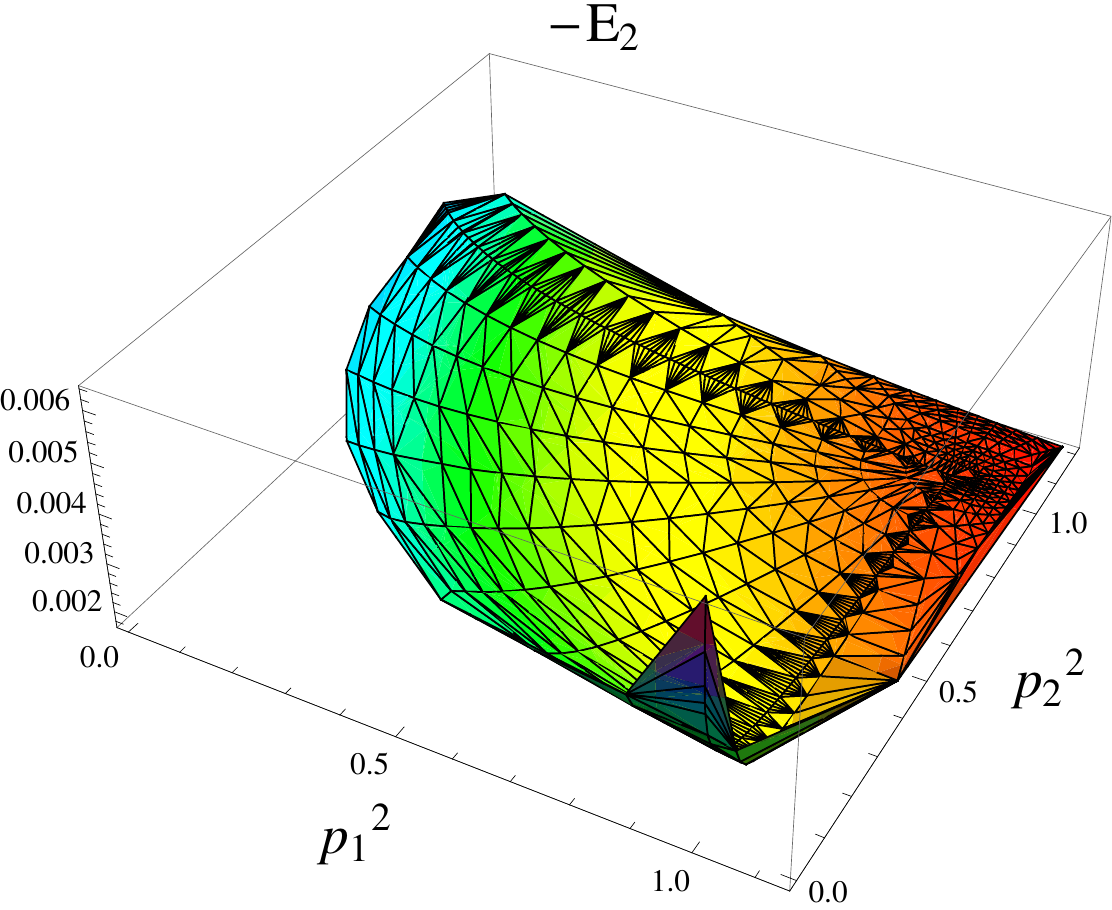,width=\linewidth}
\end{minipage}\\
\begin{minipage}[c]{0.45\linewidth}
\epsfig{file=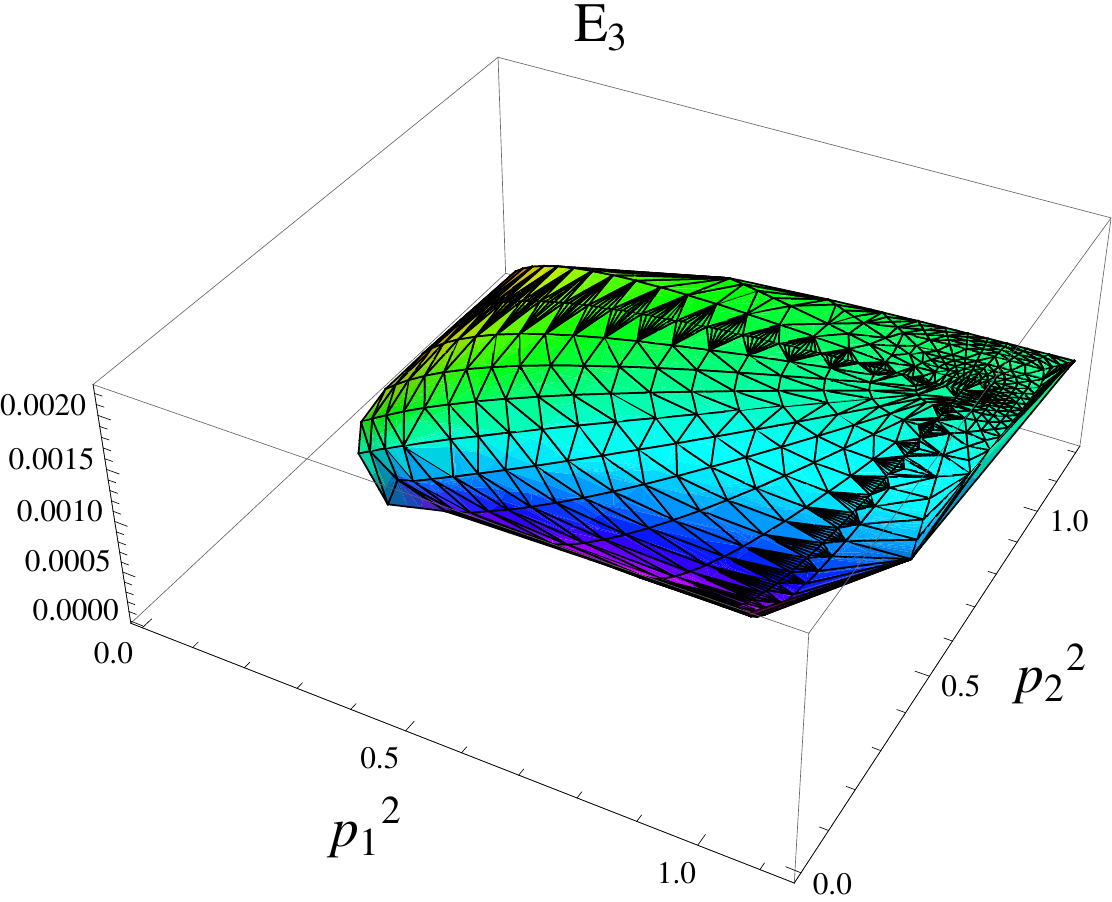,width=\linewidth}
\end{minipage}
\begin{minipage}[c]{0.45\linewidth}
\epsfig{file=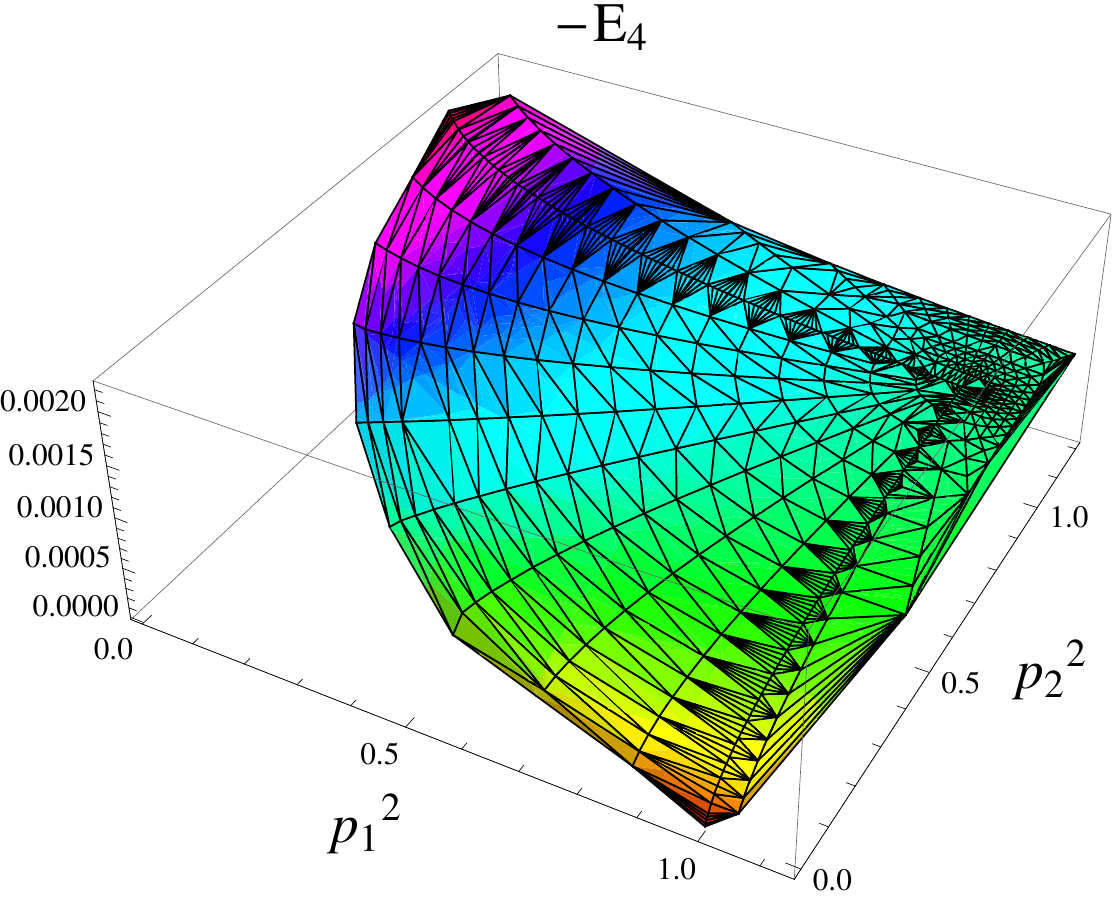,width=\linewidth}
\end{minipage}\\
\begin{minipage}[c]{0.45\linewidth}
\epsfig{file=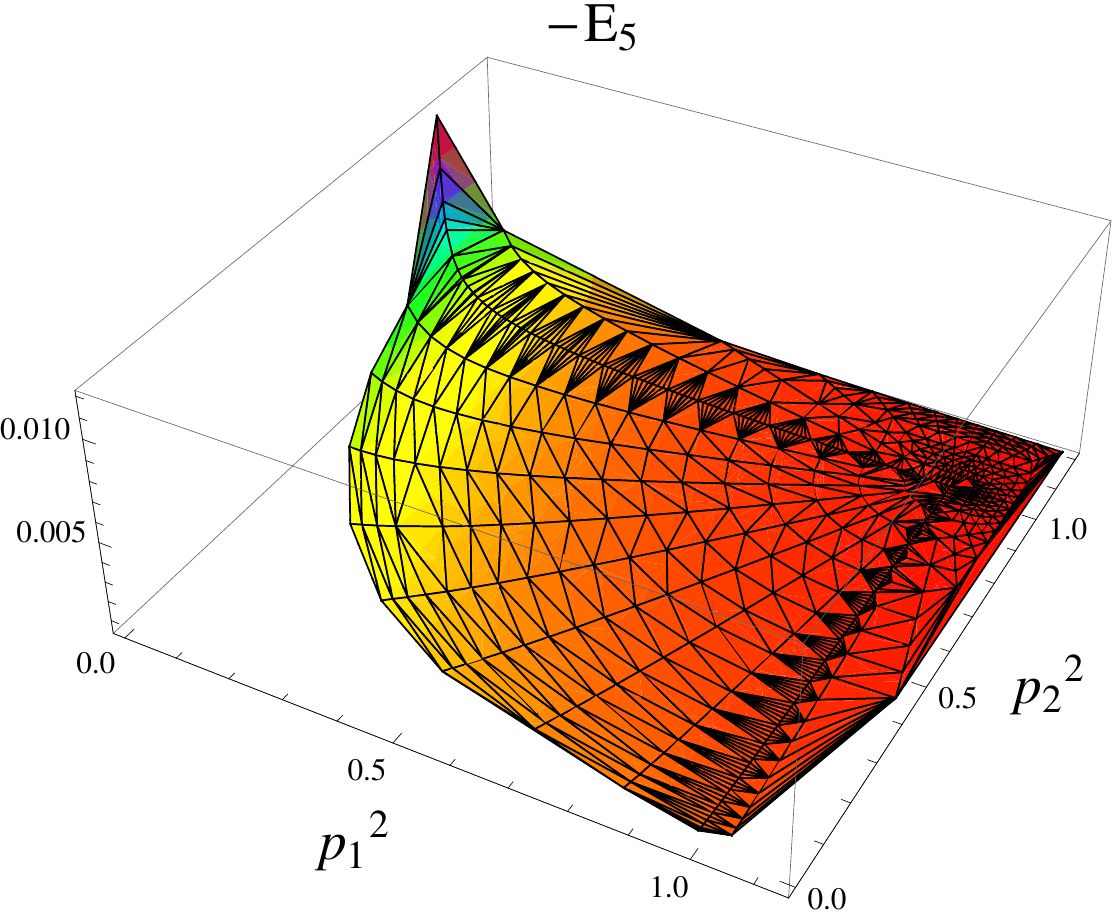 ,width=\linewidth}
\end{minipage}
\begin{minipage}[c]{0.45\linewidth}
\caption{\label{fig:resultstau15}The dressing functions corresponding to the tensors $\tau^1_{\mu \nu \rho}(p_1,p_2)
$ - $\tau^5_{\mu \nu \rho}(p_1,p_2)$ ($E_2$, $E_4$ and $E_5$ negative).}
\end{minipage}
\end{figure*}

\begin{figure*}
\begin{minipage}[c]{0.45\linewidth}
\epsfig{file=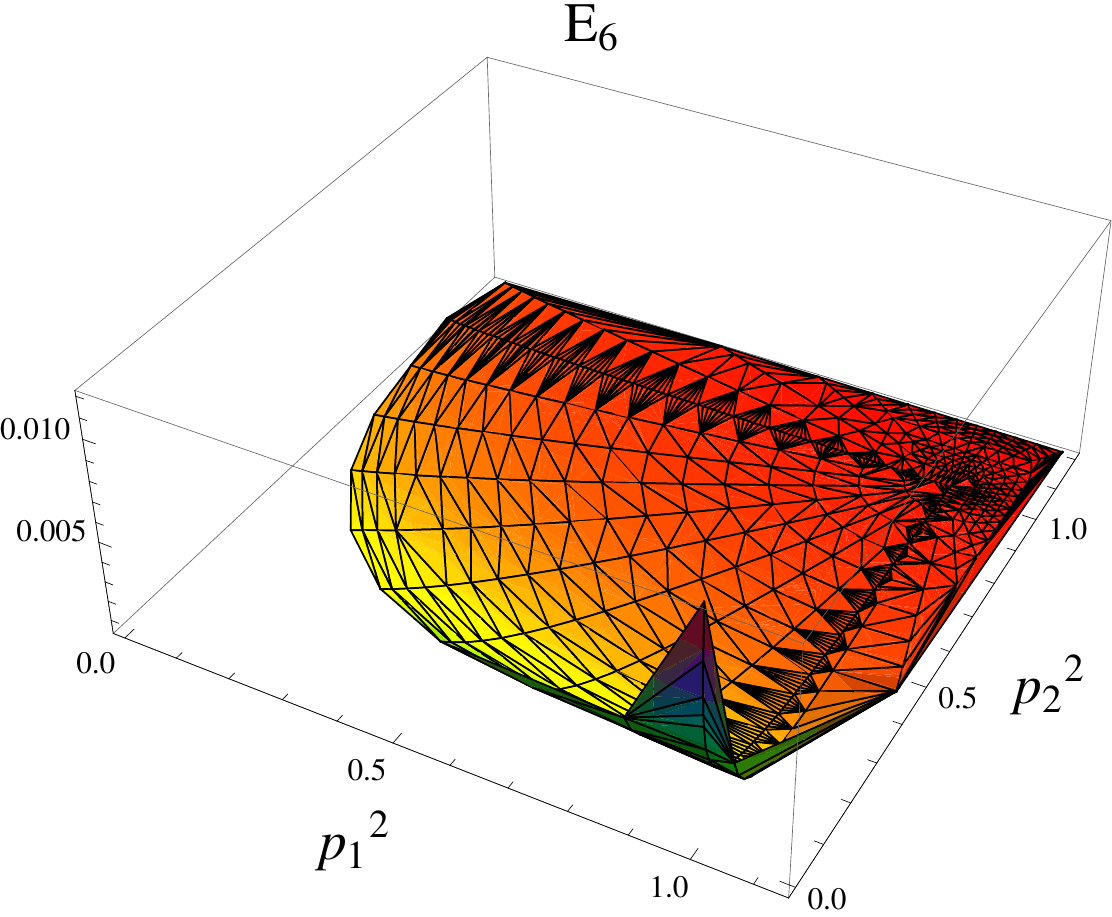,width=\linewidth}
\end{minipage}
\begin{minipage}[c]{0.45\linewidth}
\epsfig{file=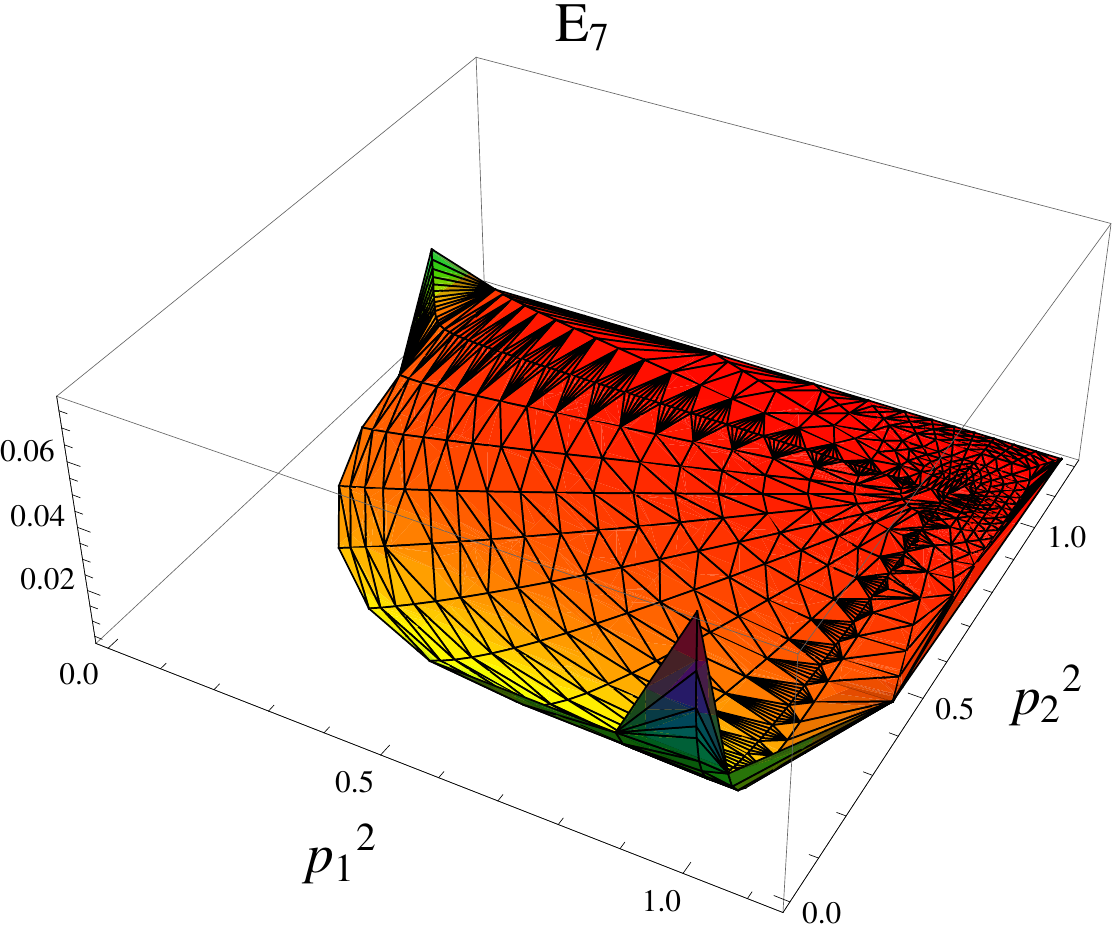,width=\linewidth}
\end{minipage}\\
\begin{minipage}[c]{0.45\linewidth}
\epsfig{file=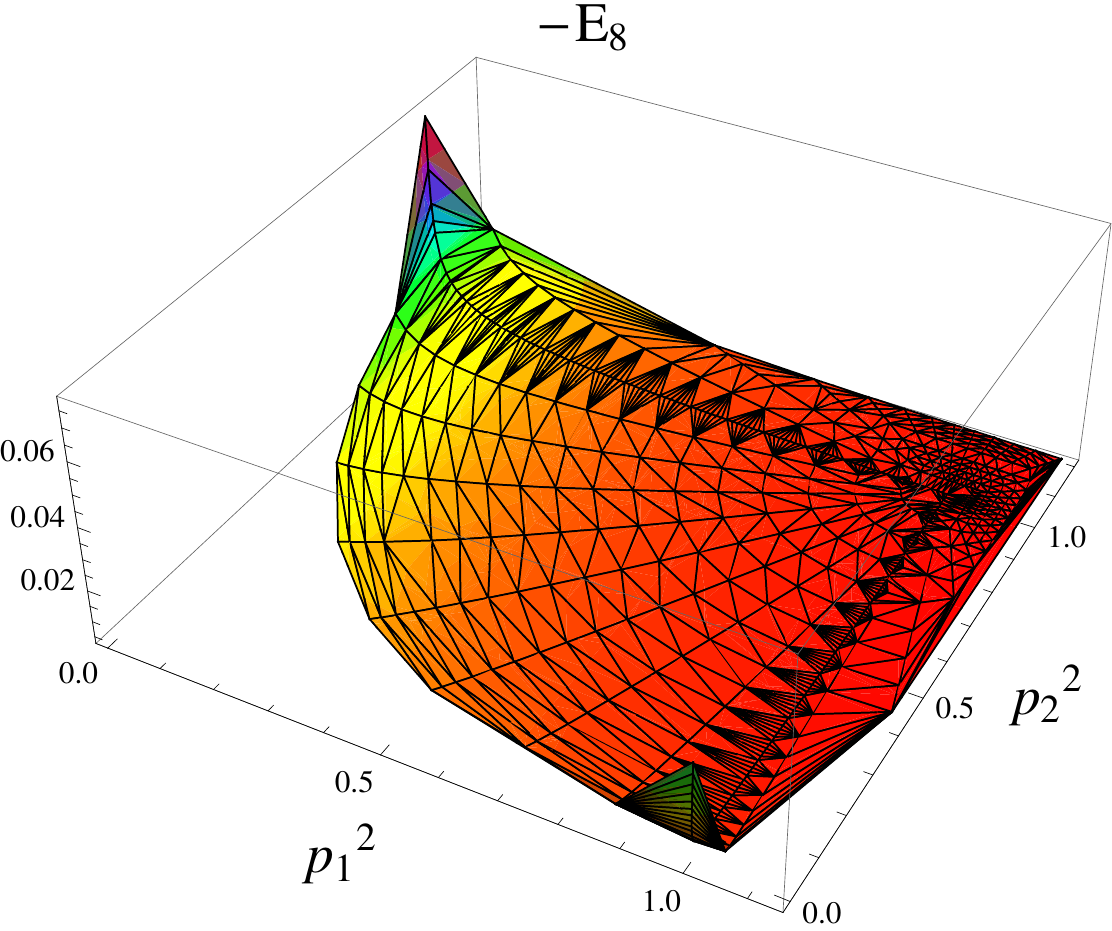,width=\linewidth}
\end{minipage}
\begin{minipage}[c]{0.45\linewidth}
\epsfig{file=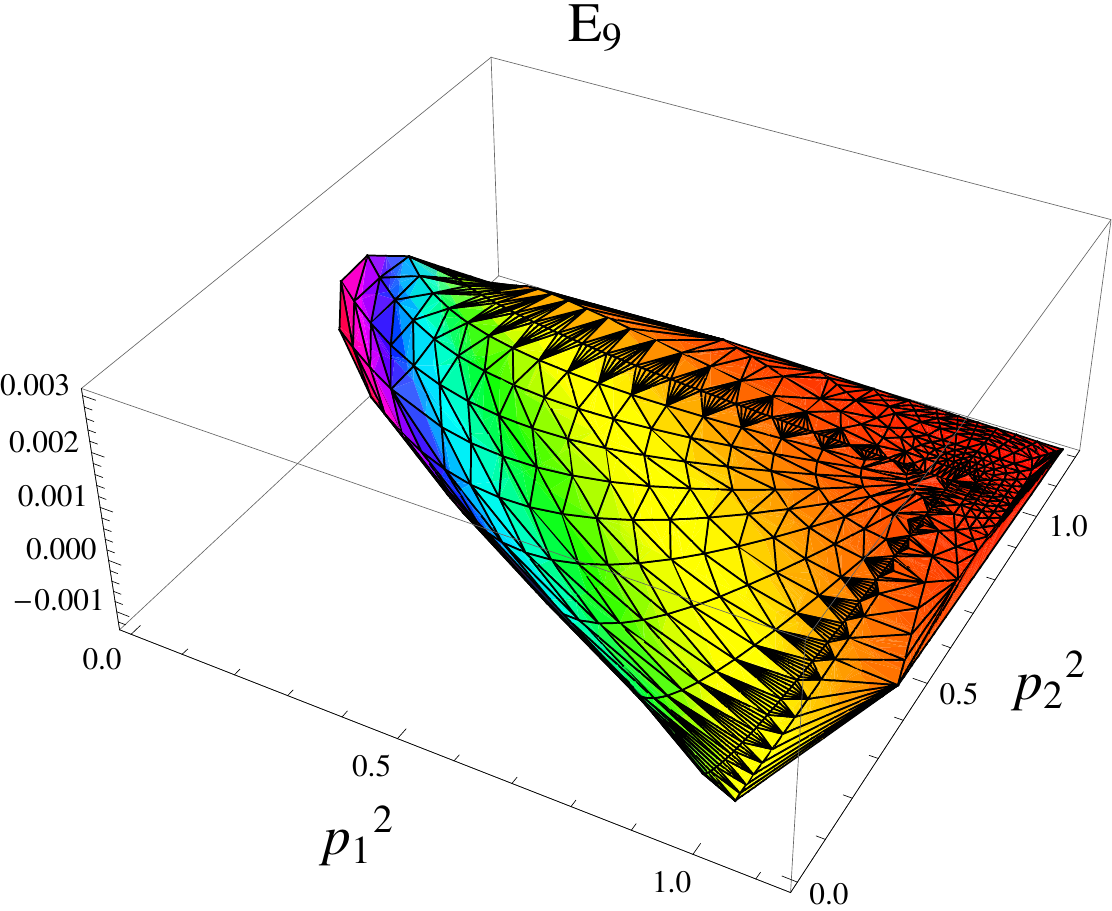,width=\linewidth}
\end{minipage}\\
\begin{minipage}[c]{0.45\linewidth}
\epsfig{file=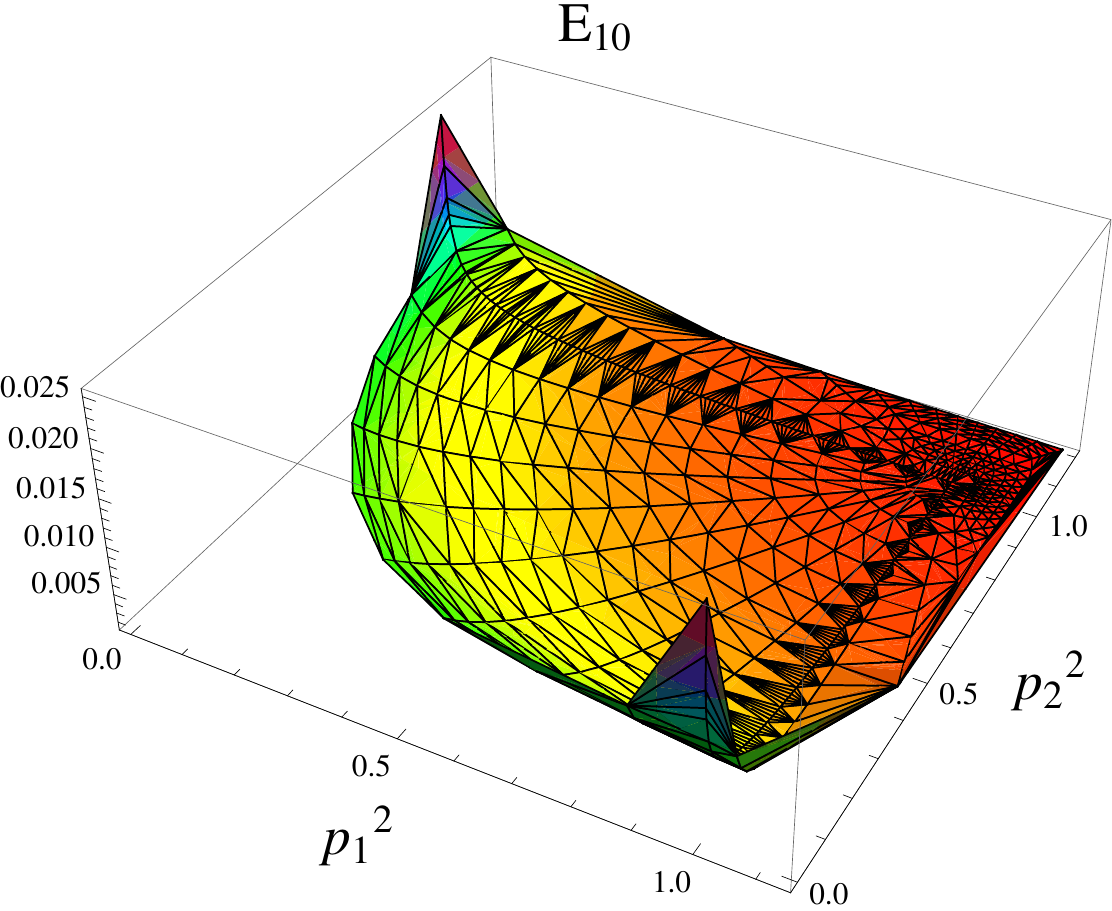,width=\linewidth}
\end{minipage}
\begin{minipage}[c]{0.45\linewidth}
\caption{\label{fig:resultstau610}The dressing functions corresponding to the tensors $\tau^6_{\mu \nu \rho}(p_1,p_2)$
 - $\tau^{10}_{\mu \nu \rho}(p_1,p_2)$
 ($E_8$ negative).}
\end{minipage}
\end{figure*}

The results for the ten dressing functions are given in figs. \ref{fig:resultstau15} and \ref{fig:resultstau610}.
Although they only mildly change near the symmetric point,
$p_1^2=p_2^2=p_3^2$, there is a strong dependence in the region near the boundary corresponding to the collinear kinematics $p_1 \sim p_2$ and in particular when approaching the singularities at the asymmetric points $p_i=0$. In contrast, previously used parametrizations of the vertex took only into account its scale dependence and therefore a constant function in the corresponding figures.
To study the behavior near the asymmetric points in more detail we
calculated the line starting at the symmetric point as shown in \fref{fig:D1001Small}. It has the advantage of fewer terms necessary to be calculated, because $y=1$, and we can get as far as $p_1^2/p_3^2=10^{-8}$. Interestingly the power law behavior sets in for
rather low ratios of $p_1^2/p_3^3$, 
but in the region below $10^{-3}$ the power laws behavior can clearly be recognized.
At each asymmetric point five dressing functions are diverging
whereas the remaining ones simply become constant or even vanish.

We can get values for the exponents by calculating the approach to the asymmetric points. 
Fitting the exponents according to a power law and expressing the result in terms of $\ka=0.5953$ gives the values reported in table \ref{tab:IR-Exps-Num} with rather mild kinematic singularities with exponent
$1-2\kappa$. Note that the only contributions that really diverge in the case of small $p_1^2$ stem from $E_7 \tau_7$ and $E_{10} \tau_{10}$, since the other divergent scalars
are rendered effectively vanishing when multiplied by the corresponding tensors. However, as in Landau gauge the gluon propagator is purely transverse, the longitudinal
parts of vertices do not contribute in DSEs \cite{Fischer:2008uz}. The tensors of $E_7$ as well as of the part of $E_{10}$ that is not suppressed by $p_1$ are longitudinal with
respect to $p_1$ and $p_3$ respectively. Therefore we do not find divergent transverse parts of this diagram that contribute in any DSE. This is in agreement with our
previous power counting analysis \cite{Alkofer:2008jy}: If the external legs of the ghost triangle in \fref{fig:triangle} are projected transversely the canonical dimensions of all three ghost-gluon vertices scale with the soft momentum, so that the expected IR exponent is
\begin{align}
\delta_{3g}^{\Delta}=\frac{d}{2}+\frac{1}{2}+2(\delta_{gh}-1)+2(\delta_{gg}+\frac1{2})=\frac{3}{2}-2\ka
\end{align}
in contrast to the non-projected case, where we had $1-2\ka$. This means that in the transversely projected case the bare vertex is leading in the soft gluon limit, $\delta_{3g,t}^{gl}=0$. However, if the vertex is contracted with some tensor, cancelations may occur so that the ghost-triangle is dominant.
We show in subsec. \ref{ssec:Lattice} that this is indeed the case and it should be possible to observe the kinematic divergence of $E_{10}$ on the lattice due to an additional suppression of the tree-level part at the asymmetric points, when contracted with a certain tensor.
Since we do not calculate all 14 tensors, it may be possible that adding them might change this special role of $E_{10}$. We also neglected the one-loop ghost term (third diagram on the right-hand side of \fref{fig:3p-DSEs}) in the three-gluon DSE that is a leading term in the conformal limit, but one order higher in the skeleton expansion.

In table \ref{tab:IR-Exps-Num} we give the IR exponents of all ten tensor components for the physically relevant case $d=4$ and $\ka=0.5953$ \cite{Zwanziger:2001kw,Lerche:2002ep}. Interestingly the momentum dependence is not independent of $\ka$, as we checked with explicit calculations for values in the range $[0,1]$. The result can most easily be presented in a closed form when using unnormalized dressing functions: Then one can see that when the IR exponent is zero it does not become larger for lower values of $\ka$ but stays zero so that the dressing functions are constant. Due to different normalizations of the tensors this happens at different values. For example in four dimensions $E_{10}$ becomes finite for $\ka\leq 1/2$, whereas $E_1$ is still divergent. For this reason we present the complete result valid for $d=2,3,4$ and $0\leq\ka\leq1$ in terms of unnormalized tensors in table \ref{tab:IR-Exps-Num-kappa}. The additional term $(d-4)/2$ accounts for the dependence on the dimension. The extension to two and three dimensions might prove useful for instance for comparison with calculations on the lattice, where often lower-dimensional studies are performed, e.~g. \cite{Maas:2007uv,Cucchieri:2008qm}.

Let us now discuss the qualitative features of these results that should be independent of the considered truncation and their impact on physical quantities. The three-gluon vertex enters for instance in the gluon loop contribution in the DSE for the gluon propagator. For soft external momenta in the deep IR regime but hard loop momenta the kinematic divergence is directly probed. Despite their much larger support in the loop integral (two momenta can be arbitrary and only the third one has to become small) the mild soft-gluon divergence is too small that it could be relevant and is IR suppressed compared to the leading ghost loop as has been discussed in \cite{Alkofer:2008jy}. In contrast for finite momenta of the order of hadronic scales the gluon loop also becomes important. Here the kinematic divergences are not directly probed. However, the results show that they also influence the behavior for finite momenta and induce a strong dependence on the kinematics in many dressing functions. The generic feature in this case is that even for finite momenta their absolute values in the vicinity of the boundary of the Euclidean region, corresponding to the collinear kinematics, and in particular towards the asymmetric points are considerably larger than around the symmetric point. Of course, at hadronic scales our results are not complete due to neglected gluonic contributions to the three-gluon vertex that may be even of comparable size in this regime. Yet, the found pronounced kinematic dependence is not expected to be completely washed out in a more refined analysis. Correspondingly this kinematic dependence should be relevant for the quantitative behavior of the gluon propagator in the intermediate regime, which in turn has a dominant impact for the properties of bound hadronic states within Bethe-Salpeter and Fadeev approaches, see e.~g. \cite{Alkofer:2000wg}. Similar remarks hold for the quantitative influence on the quark-gluon vertex and the corresponding impact on the recently proposed confinement mechanism \cite{Alkofer:2008tt,Schwenzer:2008vt}.

\begin{table*}
\begin{tabular}{l|c|c}
Tensor & $p_1$ soft& $p_2$ soft\\
\hline
\hline
$\tau_{1_{\mu\nu\rho}}=p_{1_{\mu}}p_{1_{\nu}}p_{1_{\rho}}/p_1^2$ & $1-2\ka$ & $0$\\
\hline
$\tau_{2_{\mu\nu\rho}}=p_{2_{\mu}}p_{2_{\nu}}p_{2_{\rho}}/p_2^2$ & $0$ & $1-2\ka$\\
\hline
$\tau_{3_{\mu\nu\rho}}=(p_{1_{\mu}}p_{1_{\nu}}p_{2_{\rho}}+p_{1_{\mu}}p_{2_{\nu}}p_{1_{\rho}}+p_{2_{\mu}}p_{1_{\nu}}p_{1_{\rho}})/p_1^2$ & $2-2\ka$ & $0$\\
\hline
$\tau_{4_{\mu\nu\rho}}=(p_{1_{\mu}}p_{2_{\nu}}p_{2_{\rho}}+p_{2_{\mu}}p_{1_{\nu}}p_{2_{\rho}}+p_{2_{\mu}}p_{2_{\nu}}p_{1_{\rho}})/p_2^2$ & $0$ &$2-2\ka$ \\
\hline
$\tau_{5_{\mu\nu\rho}}=g_{\mu \nu} p_{1_{\rho}}+g_{\mu \rho} p_{1_{\nu}}+g_{\nu \rho} p_{1_{\mu}}$ & $1-2\ka$ & $0$\\
\hline
$\tau_{6_{\mu\nu\rho}}=g_{\mu \nu} p_{2_{\rho}}+g_{\mu \rho} p_{2_{\nu}}+g_{\nu \rho} p_{2_{\mu}}$ & $0$ & $1-2\ka$ \\
\hline
$\tau_{7_{\mu\nu\rho}}=p_{1_{\mu}}p_{1_{\nu}}p_{2_{\rho}}/p_1^2$ & $1-2\ka$ & $1-2\ka$ \\
\hline
$\tau_{8_{\mu\nu\rho}}=p_{1_{\mu}}p_{2_{\nu}}p_{2_{\rho}}/p_2^2$ & $1-2\ka$ & $1-2\ka$\\
\hline
$\tau_{9_{\mu\nu\rho}}=(p_{1_{\mu}}p_{2_{\nu}}(p_2-p_1)_{\rho}+(p_2-p_1)_{\mu}p_{1_{\nu}}p_{2_{\rho}})/(p_1 p_2)$ & $3/2-2\ka$ & $3/2-2\ka$\\
\hline
$\tau_{10_{\mu\nu\rho}}=g_{\nu \rho} p_{1_{\mu}}-g_{\mu \nu} p_{2_{\rho}}$ & $1-2\ka$ & $1-2\ka$
\end{tabular}
\caption{\label{tab:IR-Exps-Num}The IR exponents in terms of $\kappa$ for the normalized scalar functions corresponding to the indicated tensors of the ghost triangle when $p_1^2$ or $p_2^2$ become soft in four dimensions. Since the dependence on $\ka$ is not trivial (cf. table \ref{tab:IR-Exps-Num-kappa}), these results are only valid for $\ka\geq1/2$.
}
\end{table*}

\begin{table*}
\begin{tabular}{l|c|c|c|c}
Tensor & $p_1$ soft& $p_2$ soft\\
\hline
\hline
$\tilde{\tau}_{1_{\mu\nu\rho}}=p_{1_{\mu}}p_{1_{\nu}}p_{1_{\rho}}$ & $-2\ka+\frac{d-4}{2}$ & $0$\\
\hline
$\tilde{\tau}_{2_{\mu\nu\rho}}=p_{2_{\mu}}p_{2_{\nu}}p_{2_{\rho}}$ & $0$ & $-2\ka+\frac{d-4}{2}$\\
\hline
$\tilde{\tau}_{3_{\mu\nu\rho}}=p_{1_{\mu}}p_{1_{\nu}}p_{2_{\rho}}+p_{1_{\mu}}p_{2_{\nu}}p_{1_{\rho}}+p_{2_{\mu}}p_{1_{\nu}}p_{1_{\rho}}$ & $\min(0,1-2\ka+\frac{d-4}{2})$ & $0$\\
\hline
$\tilde{\tau}_{4_{\mu\nu\rho}}=p_{1_{\mu}}p_{2_{\nu}}p_{2_{\rho}}+p_{2_{\mu}}p_{1_{\nu}}p_{2_{\rho}}+p_{2_{\mu}}p_{2_{\nu}}p_{1_{\rho}}$ & $0$ &$\min(0,1-2\ka+\frac{d-4}{2})$ \\
\hline
$\tilde{\tau}_{5_{\mu\nu\rho}}=g_{\mu \nu} p_{1_{\rho}}+g_{\mu \rho} p_{1_{\nu}}+g_{\nu \rho} p_{1_{\mu}}$ & $\min(0,1-2\ka+\frac{d-4}{2})$ & $0$\\
\hline
$\tilde{\tau}_{6_{\mu\nu\rho}}=g_{\mu \nu} p_{2_{\rho}}+g_{\mu \rho} p_{2_{\nu}}+g_{\nu \rho} p_{2_{\mu}}$ & $0$ & $\min(0,1-2\ka+\frac{d-4}{2})$ \\
\hline
$\tilde{\tau}_{7_{\mu\nu\rho}}=p_{1_{\mu}}p_{1_{\nu}}p_{2_{\rho}}$ & $-2\ka+\frac{d-4}{2}$ & $\min(0,1-2\ka+\frac{d-4}{2})$ \\
\hline
$\tilde{\tau}_{8_{\mu\nu\rho}}=p_{1_{\mu}}p_{2_{\nu}}p_{2_{\rho}}$ & $\min(0,1-2\ka+\frac{d-4}{2})$ & $-2\ka+\frac{d-4}{2}$\\
\hline
$\tilde{\tau}_{9_{\mu\nu\rho}}=p_{1_{\mu}}p_{2_{\nu}}(p_2-p_1)_{\rho}+(p_2-p_1)_{\mu}p_{1_{\nu}}p_{2_{\rho}}$ & $\min(0,1-2\ka+\frac{d-4}{2})$ & $\min(0,1-2\ka+\frac{d-4}{2})$\\
\hline
$\tilde{\tau}_{10_{\mu\nu\rho}}=g_{\nu \rho} p_{1_{\mu}}-g_{\mu \nu} p_{2_{\rho}}$ & $\min(0,1-2\ka+\frac{d-4}{2})$ & $\min(0,1-2\ka+\frac{d-4}{2})$
\end{tabular}
\caption{\label{tab:IR-Exps-Num-kappa}The IR exponents for the unnormalized scalar functions in terms of $\ka$ corresponding to the indicated tensors of the ghost triangle when $p_1^2$ or $p_2^2$ become soft. In two and three dimensions additional contributions arise. The choice of unnormalized tensors is only for an easier and more intuitive representation of the results. For comparison we give the values of $\ka$ in two, three and four dimensions that are currently believed to be the most reliable ones: $0.2$, $0.39\ldots$ and $0.59\dots$ respectively \cite{Zwanziger:2001kw,Lerche:2002ep,Maas:2004se}.
}
\end{table*}

\fig{ht}{fig:D1001Small}{D1001SmallBW,width=0.8\linewidth}{Approaching the asymmetric point $(0,1)$, i.~e. $p_1^2\rightarrow0$. Five scalars diverge, three are constant and two vanish. From top to bottom at the left side the scalars are: $E_8$, $E_{10}$, $E_5$, $E_7$, $E_1$, $E_2$, $E_4$, $E_6$, $E_9$, $E_3$. The corresponding case for vanishing $p_2^2$ can be inferred from the Bose symmetry of the three-gluon vertex.}

\subsection{Three-Gluon Vertex on the Lattice}
\label{ssec:Lattice}

Since this article is the first that gives qualitative numerical results for kinematic IR singularities it is of course of interest to investigate the possibilities of other
methods to confirm their existence. One natural possibility are lattice Monte-Carlo simulations. In this subsection we want to compare our results with recent lattice studies and investigate, if there exists a possibility to observe kinematic singularities on the lattice.

In general a comparison can only be done for the four transverse components of the three-gluon vertex, i.e.\ the tensors
$\tau_1$, $\tau_2$, $\tau_7$ and $\tau_8$ will give no contributions to quantities calculated on the lattice. Furthermore, a divergence would not manifest itself in the same way as in the continuum, because on a finite lattice results are naturally always finite. Thus, calculations can be done directly at an asymmetric point and singularities can only be seen in the volume dependence of the investigated quantity. Although there have been calculations directly at an asymmetric point \cite{Cucchieri:2006tf,Maas:2007uv,Cucchieri:2008qm}, they have not found any indication of a singularity. The quantity used in refs. \cite{Cucchieri:2006tf,Cucchieri:2008qm} was a projection of the full three-gluon vertex to the tree-level expression normalized appropriately:
\begin{align}\label{eq:lattCont}
G = \frac{\Gamma^{(0)}_{\mu \nu \rho}(p_i) D_{\mu \alpha}(p_1) D_{\nu \beta}(p_2) D_{\rho \gamma}(p_3) \Gamma_{\alpha \beta \gamma}(p_i)}{\Gamma^{(0)}_{\mu \nu \rho}(p_i) D_{ \mu \alpha}(p_1) D_{\nu \beta}(p_2) D_{\rho \gamma}(p_3) \Gamma^{(0)}_{\alpha \beta \gamma}(p_i)}.
\end{align}
Gluon propagators are denoted by $D$ and the dressed and bare three-gluon vertices by $\Gamma$ and $\Gamma^{(0)}$, respectively.
Our results indicate that no divergent contribution from the ghost triangle can be expected for $p_2^2=p_3^2$ and $p_1^2\rightarrow 0$ and there is no contradiction between lattice studies and our results from the DSE approach as far as the three-gluon vertex is concerned. It can be shown analytically that the only surviving scalar in this limit is $E_6$, which is finite at $p_1^2=0$. The corresponding numerical result is shown in \fref{fig:Latticep2Small}.

Projections with other tensors may yield different results. To investigate this possibility we resort to the Ball-Chiu basis \cite{Ball:1980ax}, which explicitly provides four transverse tensors. These are a natural basis for future studies of the full three-gluon vertex on the lattice, where only the transverse part of the vertex is accessible:
\begin{align}
	&F^3_{\mu\nu\rho}(p_1,p_2,p_3)=
		(\delta_{\mu \nu} p_1 \cdot p_2 - p_{1_{\nu}}p_{2_{\mu}}){\cal B}^3_{\rho},\\
	&F^1_{\mu\nu\rho}(p_2,p_3,p_1)=
		(\delta_{\nu \rho} p_2 \cdot p_3 - p_{2_{\rho}}p_{3_{\nu}}){\cal B}^1_{\mu},\\
	&F^2_{\mu\nu\rho}(p_3,p_1,p_2)=
		(\delta_{\mu \rho} p_3 \cdot p_1 - p_{3_{\mu}}p_{1_{\rho}}){\cal B}^2_{\nu},\\
	&H_{\mu\nu\rho}(p_1,p_2,p_3)=
		(-\delta_{\mu \nu}{\cal B}^3_{\rho}
		-\delta_{\nu \rho}{\cal B}^1_{\mu}
		-\delta_{\mu \rho}{\cal B}^2_{\nu} +\nonumber\\
	&\qquad+p_{1_{\rho}}p_{2_{\mu}}p_{3_{\nu}}-p_{1_{\nu}}p_{2_{\rho}}p_{3_{\mu}}),
\end{align}
with $B_\mu^i=\epsilon_{ijk} p_{j_\mu} p_k \cdot p_i$.
The tensor $H_{\mu\nu\rho}$ is the only one that depends on $p_1$ in every term, so that it scales like $(p_1^2)^{1/2}$ in the corresponding limit.
The calculated contractions are:
\begin{align}\label{eq:Contractions}
\bar{F}^i(p_1^2,p_2^2,p_3^2)=\frac{1}{p_1^2}F^i_{\mu\nu\rho}(p_1,p_2,p_3)P_{\mu\alpha}P_{\nu\beta}P_{\rho\gamma}\Gamma_{\alpha\beta\gamma}^\Delta,\nnnl
\bar{H}(p_1^2,p_2^2,p_3^2)=\frac{1}{p_1^2}H_{\mu\nu\rho}(p_1,p_2,p_3)P_{\mu\alpha}P_{\nu\beta}P_{\rho\gamma}\Gamma_{\alpha\beta\gamma}^\Delta,
\end{align}
where $P_{\mu\nu}$ is the standard transverse projector and $\Gamma_{\alpha\beta\gamma}^\Delta$ the ghost triangle part of the three-gluon vertex.
The results are given in \fref{fig:Lattice-Ball-Chiu}. The normalization $1/p_1^2$ was chosen to show the qualitative different behavior of the $F^i$ and the $H$ quantities. Analytic considerations show that directly at the asymmetric point, i.~e. $p_1^2=0$ and $p_2^2=p_3^2$, the corresponding values are
\begin{align}
\bar{F}^1(0,p_2^2,p_2^2)&=3 (p_2^2)^2E_6(0,p_2^2,p_2^2),\\
\bar{F}^2(0,p_2^2,p_2^2)&=\bar{F}^3(0,p_2^2,p_2^2)=E_6 (p_2^2)^2(0,p_2^2,p_2^2),\\
\bar{H}(0,p_2^2,p_2^2)&=2 E_{10}  p_2^2(0,p_2^2,p_2^2).
\end{align}
Thus the $\bar{F}^i$ are finite and $\bar{H}$ diverges like $(p_1^2)^{1-2\ka}$.
Normally one would expect that the tree-level gives the dominant contribution, since it does not scale in the soft gluon limit, whereas the transverse part of $E_{10}$ is suppressed.
However, it turns out that in the case $p_2^2=p_3^2$ the contraction of the tree-level with the $H$-tensor yields an additional suppression, i.~e. although one expects that $H\,P\,P\,P\,\Gamma^{(0)}$ scales like $(p_1^2)^{1/2}$, because of the $p_1$-dependence of the $H$-tensor, it goes like $(p_1^2)^1$. With normalization as given above this yields an IR finite contribution from the tree-level $\Gamma^{(0)}$ and the ghost-triangle $\Gamma^{\Delta}$ is the leading part in the contraction of the full vertex $\Gamma=\Gamma^{(0)}+\Gamma^{\Delta}$ with the $H$-tensor giving a power law with a non-integer exponent. Thus it should be possible to verify the existence of kinematical singularities also on the lattice by studying the full vertex. However, we want to stress that only for $p_2^2=p_3^2$ the leading part obeys a power law with an exponent that depends on $\ka$ due to an additional suppression of the $H$-tensor, whereas otherwise the tree-level is dominant.

\fig{ht}{fig:Latticep2Small}{Latticep1Small,width=0.8\linewidth}{The normalized contraction of the ghost triangle of the three-gluon vertex with three propagators and a tree-level vertex as given in \eref{eq:lattCont}. The value clearly approaches a constant for $p_1^2\rightarrow 0$.}

\fig{ht}{fig:Lattice-Ball-Chiu}{Lattice-Ball-Chiu,width=0.8\linewidth}{The contractions of the ghost triangle with the transverse tensors of the Ball-Chiu basis as given in \eref{eq:Contractions} and the contraction of the tree-level tensor with the $H$ tensor along the line from the symmetric to the asymmetric point $p_1^2=0$. The values for $\bar{F}_2$ and $\bar{F}_3$ coincide.}

\section{Tensor Components of the Ghost-Gluon Vertex in the Infrared}\label{sec:resultsGhGV}

The ghost-gluon vertex can in principle be treated along the same lines as the three-gluon vertex. One might even suspect that the actual calculation is simpler, because the full vertex only has two possible tensors,
\begin{align}
\label{eq:GhGTensors}
&\Gamma_\mu(p_{gh}^{(in)},p_{gh};p_{gl})=\nnnl
&=i\, A((p_{gh}^{(in)})^2,p_{gh}^2,p_{gl}^2)\, p_{{gh}_\mu}+i\, B((p_{gh}^{(in)})^2,p_{gh}^2,p_{gl}^2)\, p_{{gl}_\mu},
\end{align} 
where $p_{gh}$/$p_{gh}^{(in)}$ is the momentum of the outgoing/incoming ghost and $p_{gl}$ that of the gluon.
However, there are some additional complications that are due to the uniform IR exponent of the ghost-gluon vertex being $0$. Thereby the combinations of propagator IR exponents can combine to $0$ in the Gamma functions of \eref{eq:I-final}. This complicates the analytic continuation, because the limit has to be taken carefully in which case 
the divergent Gamma functions cancel each other to leave a finite result. The corresponding formulae that have to be used are given in appendix \ref{sec:anal-cont-F4}.

From the two existing DSEs for the ghost-gluon vertex we choose the seemingly more complicated one given in \fref{fig:3p-DSEs}, where the bare vertices are attached to the external gluon leg. In this case the leading diagrams are the ghost-ghost-gluon triangle and the ghost loop (diagrams 4 and 5) as well as the bare vertex itself. The first order of the skeleton expansion consists of the first diagram only, since the ghost-ghost scattering kernel in the second is one-particle irreducible. The alternative version of the DSE features a leading diagram involving a dressed three-gluon vertex, whereas here only dressed ghost-gluon vertices appear and we employ again bare vertices. We denote the contributions to the dressing functions $A$ and $B$ from the ghost-ghost-gluon triangle by $A^\Delta$ and $B^\Delta$ respectively.

\begin{figure}
\begin{minipage}[c]{0.9\linewidth}
\resizebox{\linewidth}{!}{\epsfig{file=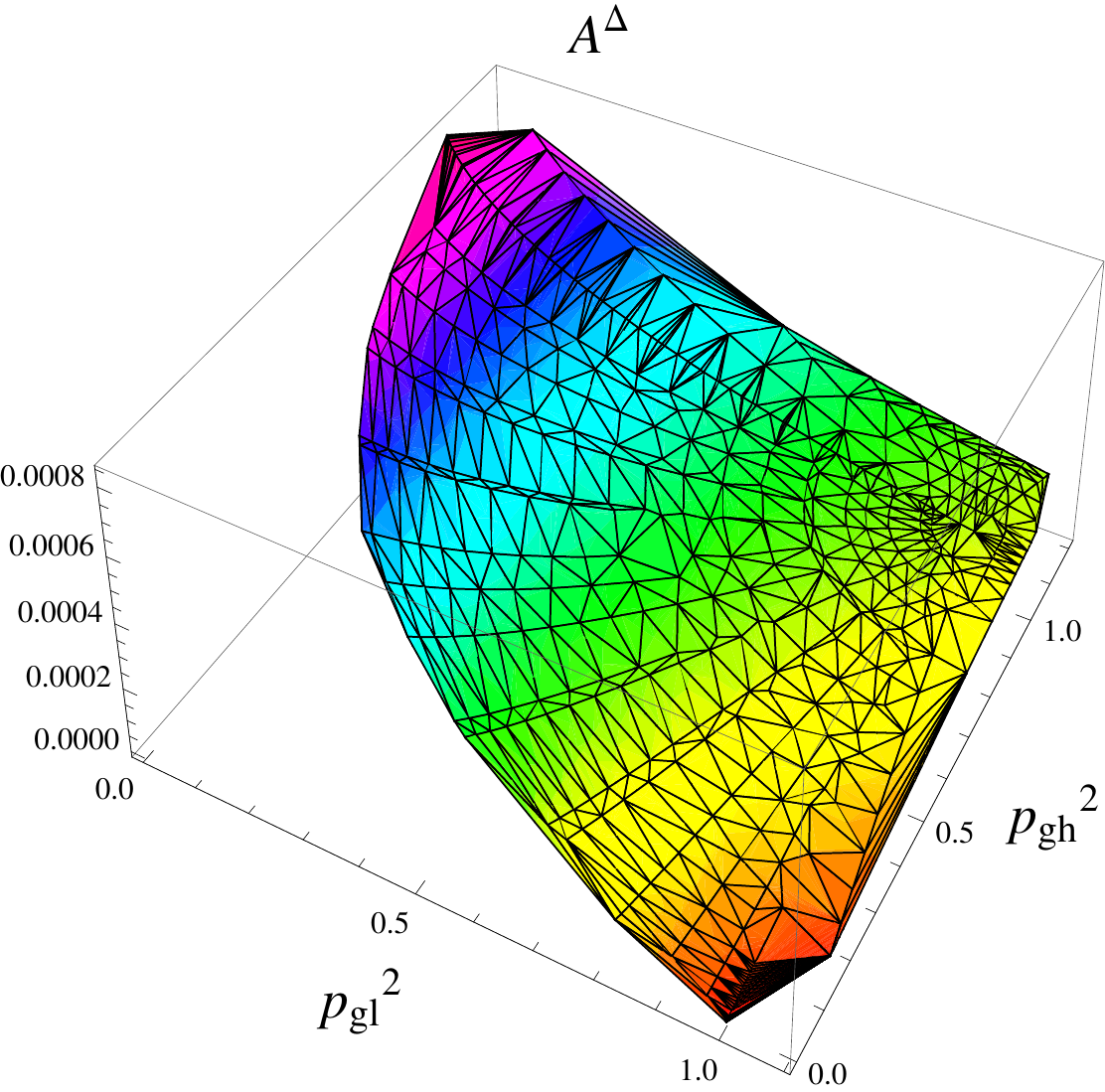,width=\linewidth}}
\end{minipage}
\begin{minipage}[c]{0.9\linewidth}
\resizebox{\linewidth}{!}{\epsfig{file=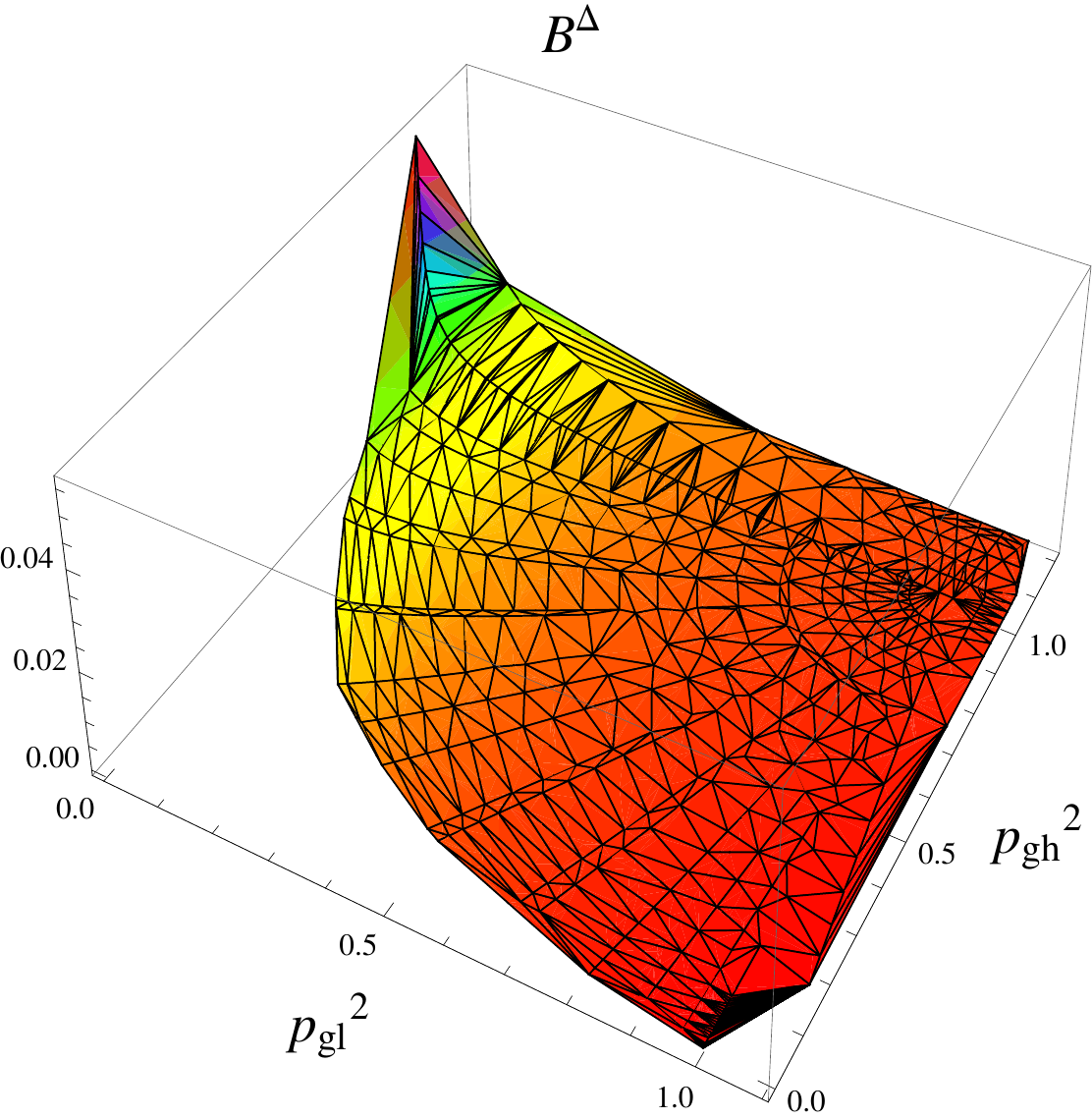,width=\linewidth}}
\end{minipage}

\caption{\label{fig:resultsGhGV} The dressing function contributions from the ghost-ghost-gluon triangle corresponding to the tensors $p_{gh_\mu}$ (ghost leg) and $p_{gl_\mu}$ (gluon leg).}
\end{figure}

\begin{figure*}
\epsfig{file=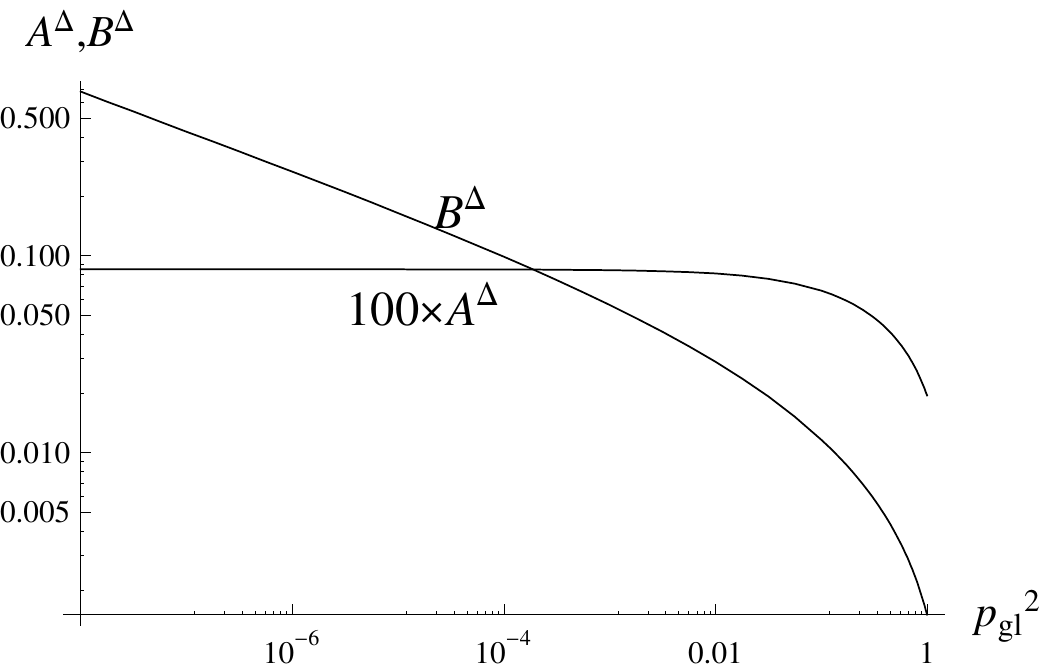,width=0.45\linewidth}
\epsfig{file=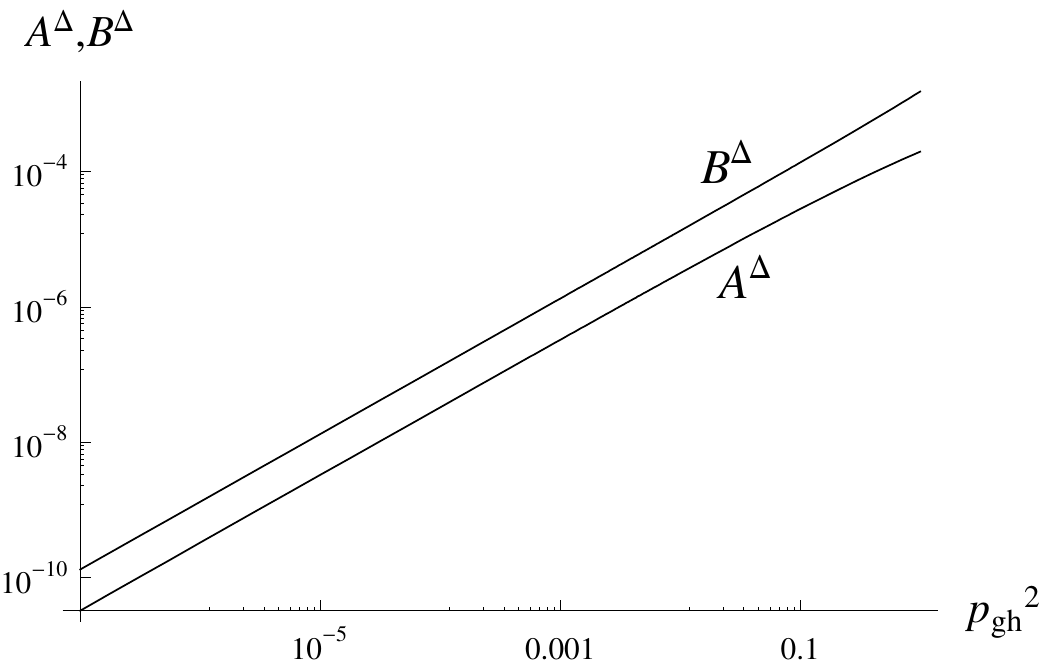,width=0.45\linewidth}

\caption{\label{fig:FSmall} Approaching the asymmetric points $p_{gl} \to 0$ (left) and $p_{gh} \to 0$ (right). In the former case the ghost-ghost-gluon triangle dressing function of the gluon leg, $B^\Delta$, is divergent and that of the ghost leg, $A^\Delta$, is constant whereas in the latter all scalars vanish.}
\end{figure*}

The results for the ghost-ghost-gluon triangle in the Euclidean region are shown in \fref{fig:resultsGhGV}. The approach to the asymmetric points is plotted in \fref{fig:FSmall}. Here we have to investigate both limits of the gluon and the ghost leg momentum going to zero because there is no Bose symmetry as in the case of the three-gluon vertex. Numerical values for the exponents are given in table \ref{tab:IR-Exps-Num-Triangle} for the ghost-ghost-gluon triangle and in table \ref{tab:IR-Exps-Num-GhGV} for the overall vertex. In four dimensions only the longitudinal dressing function $B$ exposes a divergence when the gluon momentum goes to zero. Therefore the existence of kinematical divergences in the ghost-gluon vertex cannot be investigated on the lattice. Furthermore the divergent dressing function is multiplied with a tensor that also scales with the gluon momentum, rendering this contribution to the vertex IR vanishing in this limit.
Our results agree with the statement of Taylor \cite{Taylor:1971ff} that for vanishing ghost momentum the ghost-gluon vertex becomes bare, i.~e. the IR exponent is $0$, because the loop corrections in
the DSE are indeed subleading compared to the bare vertex.
The dependence on $\ka$ is similar to that of the
ghost triangle, i.~e. the IR exponent can also become $0$ for values of $\ka$ below a certain value determined by the dimension. Whereas in four dimensions the transverse dressing function $A^\Delta$ is always finite for values of $\ka$ between 0 and 1 it turns out that there is indeed a dependence on $\ka$ that is revealed in lower dimensions. However, the relevant values of $\ka$ are well below this regions, see table \ref{tab:IR-Exps-Num-GhGV}.

\begin{table}
\begin{tabular}{l|c|c}
Dressing function & $p_{gl}\to 0$ & $p_{gh}\to 0$ \\
\hline
\hline
$A^\Delta$ & $\min(0,2-2\ka+\frac{d-4}{2})$  & $ 1$ \\
\hline
$B^\Delta$ & $\min(0,1-2\ka+\frac{d-4}{2})$ & $ 1$ \\
\hline
tree-level & $-$ & $0$ \\
\end{tabular}
\caption{\label{tab:IR-Exps-Num-Triangle}The IR exponents of the ghost-ghost-gluon triangle and the tree-level when $p_{gl}$ or $p_{gh}$ become soft. }
\end{table}

\begin{table}
\begin{tabular}{l|c|c}
Dress. fct. $\times$ tensor & $p_{gl}\to 0$ & $p_{gh}\to 0$ \\
\hline
\hline
$ A\,p_{{gh}_\mu}$ & $\min(0,2-2\ka+\frac{d-4}{2})$  & $1/2$ \\
\hline
$ B\,p_{{gl}_\mu}$ & $\min(\frac1{2},\frac{3}{2}-2\ka+\frac{d-4}{2})$ & $ 1$ \\
\end{tabular}
\caption{\label{tab:IR-Exps-Num-GhGV}The overall IR exponents of the ghost-gluon vertex, i.e. ghost-ghost-gluon triangle and the bare vertex including the canonical momenta. The dominant part of the transverse dressing function $A$ is then given by the bare vertex.}
\end{table}

\section{Summary}

We presented an analytic solution for the IR leading parts of the three-gluon and ghost-gluon vertices, which are given by the ghost triangle and the ghost-ghost-gluon
triangle. We found the expected uniform IR behavior, i.e. an IR exponent of $-3\ka$ for the three-gluon vertex and $0$ for the ghost-gluon vertex, but also confirmed the
existence of kinematic singularities in the case that only one momentum goes to zero as predicted in ref. \cite{Alkofer:2008jy}. As detailed above the non-trivial momentum dependence in the Euclidean region may have a quantitative impact on other Green functions.

For our calculations we combined known
results for the massless three-point integral and analytic continuations of Appell's function $F_4$. Using bare ghost-gluon vertices we found that for the three-gluon vertex
five of ten dressing functions are divergent for one momentum going to zero. Although in this limit all transverse parts are additionally suppressed, it should be possible to confirm the existence of kinematic divergencies on the lattice, since the otherwise leading tree-level tensor is additionally suppressed at the asymmetric point.

Also for the ghost-gluon vertex there is a divergent dressing function that is suppressed by the corresponding tensor. Thus the vertex itself shows no kinematic singularities. We determined the dependence of the dressing functions on $\ka$ in the region $0<\ka<1$ for two, three and four dimensions.

\section*{Acknowledgments}

We would like to thank Christian Fischer, Felipe Llanes-Estrada and Selym Villalba Chavez for fruitful discussions. We are grateful to Axel Maas for a detailed explanation of his lattice results and a critical reading of the manuscript. This work was supported in part by grant 
M979-N16 of the Austrian Research Foundation FWF.

\appendix

\section{Analytic Continuation of Appell's Function $F_4$ into the Euclidean Momentum Region}\label{sec:anal-cont-F4}

In this appendix we derive the analytically continued expression of Appell's function $F_4$ that was used in the calculations for three-point integrals.
The original derivation was performed in ref.\ \cite{Exton:1994de}. We noted that this reference contained a few typos which we will correct in the following. 
In addition we discuss an appropriate reformulation for the treatment of special cases of the arising series which are necessary for a numerical implementation.

\subsection{Preliminaries}

The derivation is simplified by use of the Pochhammer symbol, defined as
\equ{\label{eq:def-PochhammerWithGamma}
(a,n)=\frac{\G{a+n}}{\G{a}}.
}
Since it will be used extensively in the following derivation, we give here some standard relations. Simply by inserting one can show
\begin{equation}
\label{eq:Poch-multi}
(a,m+n)=(a,m)(a+m,n).
\end{equation}
From Legendre's duplication formula for the Gamma function a similar relation for the Pochhammer symbol follows:
\equ{\label{eq:Poch-double}
(a,2b)=2^{2b}(a/2,b)(1/2+a/2,b).
}
An important formula is the analytic continuation of the Pochhammer symbol. It is only valid for integer values of n:
\begin{equation}
\label{eq:Poch-anal-cont}
(a,-n)=\frac{(-1)^n}{(1-a,n)}.
\end{equation}
The Gaussian hypergeometric series and generalized hypergeometric series of one variable will appear, which are in general defined as
\begin{align}\label{eq:gen-hyp}
_pF_q&(a_1,\dots,a_p;b_1,\dots,b_q;z)
	\equiv\,_pF_q\left(
		\begin{matrix}
			a_1,\dots,a_p;\\
			b_1,\dots,b_q;
		\end{matrix}
		\; z \right):=\nnnl
	&:=\sum_{n=0}^\infty \frac{(a_1,n)\dots(a_p,n)}{(b_1,n) \ldots (b_q,n)}\frac{z^n}{n!}, \qquad p,q \in \mathbb{N}_0.
\end{align}
If $p=2$ and $q=1$ this is the Gaussian hypergeometric series of dimension one, where by dimension we mean the number of variables (here $z$). The $a_i$ and $b_i$ are referred to as parameters.

\begin{widetext}
\subsection{Analytic Continuation}

The first step is to rewrite the original definition of $F_4$, \eref{eq:F4-def}, in such a way that it contains the Gaussian hypergeometric series of one variable:
\begin{equation}
\label{eq:F4-with-2F1}
F_4(a,b;c,d;x,y)=\sum_{m,n=0}^\infty \frac{(a,m)(b,m)}{(c,m)}\frac{x^m}{m!}\,_2F_1(a+m,b+m;d;y).
\end{equation}
Now we use an analytic continuation of $_2F_1$ ((15.3.6) in \cite{Abramowitz:1972mf}),
\begin{align}
\label{eq:2F1-anal-cont2}
_2F_1&(a,b;c;y)=\frac{\Gamma(c)\Gamma(c-a-b)}{\Gamma(c-a)\Gamma(c-b)}\, _2F_1(a,b;a+b-c+1;1-y)+\nonumber\\
	&+\frac{\Gamma(c)\Gamma(a+b-c)}{\Gamma(a)\Gamma(b)}(1-y)^{c-a-b}\, _2F_1(c-a,c-b;c-a-b+1;1-y).
\end{align}
We insert eq. (\ref{eq:2F1-anal-cont2}) into eq. (\ref{eq:F4-with-2F1}) and treat the two appearing terms separately. The first one will lead to a series which we call $G$ according to the conventions of ref. \cite{Exton:1994de}. It is the easier one of the two emerging series:
\begin{align}
\sum_{m=0}^\infty& \frac{x^m}{m!} \frac{(a,m)(b,m)}{(c,m)} \frac{\Gamma(d)\Gamma(d-a-b-2m)}{\Gamma(d-a-m)\Gamma(d-b-m)} 
	\sum_{n=0}^\infty \frac{(1-y)^n}{n!} \frac{(a+m,n)(b+m,n)}{(a+b+2m-d+1)}=\nonumber\\
	&=\frac{\Gamma(d)\Gamma(d-a-b)}{\Gamma(d-a)\Gamma(d-b)}
		\sum_{m,n=0}^\infty \frac{x^m}{m!} \frac{(1-y)^n}{n!} \frac{(a,m+n)(b,m+n)}{(c,m)}
		\frac{(1-d+a,m)(1-d+b,m)}{(1-d+a+b,2m+n)}=\nonumber\\
	&=\frac{\Gamma(d)\Gamma(d-a-b)}{\Gamma(d-a)\Gamma(d-b)} G(a,b,1-d+a,1-d+b;1-d+a+b,c;x,1-y).
\end{align}
We have used Eqs. (\ref{eq:Poch-anal-cont}) and (\ref{eq:def-PochhammerWithGamma}) and the $G$ series is defined as
\begin{equation}\label{eq:def-G}
G(a,b,c,d;e,f;x,y):=\sum_{m,n=0}^\infty \frac{x^m}{m} \frac{y^n}{n!} \frac{(a,m+n)(b,m+n)(c,m)(d,m)}{(e,2m+n)(f,m)}.
\end{equation}

The second part is more intricate. Employing Eqs. (\ref{eq:Poch-multi}), (\ref{eq:Poch-anal-cont}) and (\ref{eq:def-PochhammerWithGamma}) we get
\begin{align}
\label{eq:H-series}
\sum_{m=0}^\infty& \frac{x^m}{m!} \frac{(a,m)(b,m)}{(c,m)}
	\frac{\Gamma(d)\Gamma(a+b-d+2m)}{\Gamma(a+m)\Gamma(b+m)}(1-y)^{d-a-b-2m}\times \nonumber\\
		&\quad\times\sum_{n=0}^\infty \frac{(1-y)^n}{n!} \frac{(d-a-m,n)(d-b-m,n)}{(1+d-a-b-2m,n)}=\nonumber\\
	&=\frac{\Gamma(d)\Gamma(a+b-d)}{\Gamma(a)\Gamma(b)} (1-y)^{d-a-b} \times \nonumber\\
	&	\quad\times\sum_{m,n=0}^\infty \left(\frac{x}{(1-y)^2}\right)^m \frac{(y-1)^n}{n!}
		\frac{(a+b-d,2m-n)(1+a-d,m)(1+b-d,m)}{(1+a-d,m-n)(1+b-d,m-n)(c,m)}.
\end{align}
In this form the series is still not convergent in the Euclidean region and we need to perform another analytic continuation. For this we rewrite eq. (\ref{eq:H-series}) into a form, which contains the generalized hypergeometric series $_4F_3$:
\begin{align}
\label{eq:H-4F3}
	&\frac{\Gamma(d)\Gamma(a+b-d)}{\Gamma(a)\Gamma(b)} (1-y)^{d-a-b} \sum_{n=0}^\infty \frac{(1-y)^n}{n!} \frac{(d-a,n)(d-b,n)}{(1-a-b+d,n)} \times\nonumber\\
	&\quad\times_4F_3\left(
			\begin{matrix}
				\frac{a}{2}+\frac{b}{2}-\frac{d}{2}-\frac{n}{2},\frac{1}{2}+\frac{a}{2}+\frac{b}{2}-\frac{d}{2}-\frac{n}{2},1+a-d,1+b-d;\\
				c,1+a-d-n,1+b-d-n;
			\end{matrix}
			\; \frac{4x}{(1-y)^2}\right).
\end{align}
Eqs. (\ref{eq:Poch-multi}), (\ref{eq:Poch-double}) and (\ref{eq:Poch-anal-cont}) were used. 
Writing the generalized hypergeometric series $_4F_3$ as a Meijer-G function $G^{k,l}_{m,n}$ (eq. (1) on page 215 in \cite{Erdelyi:1953ht}),
\begin{equation}
\label{eq:4F3asMeijer-G}
_4F_3(a,b,c,d;e,f,g;x)=\frac{\Gamma(e)\Gamma(f)\Gamma(g)}{\Gamma(a)\Gamma(b)\Gamma(c)\Gamma(d)}G^{4,1}_{4,4}\left(-\frac{1}{x}\Bigg|
	\begin{matrix}
	1,e,f,g\\
	a,b,c,d	
	\end{matrix}
	\right),
\end{equation}
we can employ its analytic continuation (eq. (5) on p. 208 in \cite{Erdelyi:1953ht}):
\begin{align}
\label{eq:Meijer-G-def}
&G^{m,n}_{p,q} \left(x\Bigg|
		\begin{matrix}
			a_1, \dots, a_p\\
			b_1, \dots, b_q
       	\end{matrix}
		\right)=\sum_{h=1}^m \frac{
		\sideset{}{'}\prod_{j=1}^{\,m} \Gamma(b_j-b_h) \prod_{j=1}^n \Gamma(1+b_h-a_j)}
		{\prod_{j=m+1}^q \Gamma(1+b_h-b_j) \prod_{j=n+1}^p \Gamma(a_j-b_h)} x^{b_h} \times \nonumber\\
	&\quad\times_pF_{q-1}\left(
		\begin{matrix}
			1+b_h-a_1, \dots, 1+b_h-a_p;\\
			1+b_h-b_1,\dots,\ast,\dots, 1+b_h-b_q;
		\end{matrix}
		\; (-1)^{p-m-n} x \right) 
		\qquad p<q \vee p=q \wedge |x|<1.
\end{align}
The primed product excludes the expression $\Gamma(0)$ (when $b=j$), and the $\ast$ stands for the one expression taken out as argument of $_pF_{q-1}$.
Inserting eq. (\ref{eq:Meijer-G-def}) into eq. (\ref{eq:4F3asMeijer-G}) shows that two of the four expected $_4F_3$ vanish due to the appearance of $\Gamma(-n)$ in the denominator which yields $0$ because $n$ is an integer. In the two functions left half-integer values of $n$ appear:
\begin{align}
&\frac{\Gamma(d)\Gamma(a+b-d)}{\Gamma(a)\Gamma(b)}(1-y)^{d-a-b}\times\nnnl
	&\quad\times \Bigg\lbrace \frac{\G{c}\G{\mhalfo}(-4x)^{\mhalf{d}-\mhalf{a}-\mhalf{b}}}{\G{c-\mhalf{a}-\mhalf{b}-\mhalf{d}}\G{\mhalf{a}+\mhalf{b}-\mhalf{d}+\mhalfo}} \sum_{m,n=0}^\infty (-x)^{\mhalf{n}}\frac{(-1)^{n+m}}{m!n!}\left(\frac{(1-y)^2}{4x}\right)^m\times \nonumber\\
		&\quad\times \frac{(1+\mhalf{a}-\mhalf{b}-\mhalf{d},\mhalf{n}-m)(1+\mhalf{b}-\mhalf{a}-\mhalf{d},\mhalf{n}-m)(\mhalf{a}+\mhalf{b}-\mhalf{d},m-\mhalf{n})}{(\mhalfo,m)(1+\mhalf{a}-\mhalf{b}-\mhalf{d},-\mhalf{n}-m)(1-\mhalf{a}+\mhalf{b}-\mhalf{d},-\mhalf{n}-m)(c-\mhalf{a}-\mhalf{b}+\mhalf{d},\mhalf{n}-m)}+\nnnl
	&+\frac{(1-y)\G{c}\G{-\mhalfo}(-4x)^{\mhalf{d}-\mhalf{a}-\mhalf{b}-\mhalfo}}{\G{c-\mhalf{a}-\mhalf{b}+\mhalf{d}-\mhalfo}\G{\mhalf{a}+\mhalf{b}-\mhalf{d}}} \sum_{m,n=0}^\infty (-x)^{\mhalf{n}}\frac{(-1)^{n+m}}{m!n!}\left(\frac{(1-y)^2}{4x}\right)^m\times \nonumber\\
		&\quad\times \frac{(1+\mhalf{a}-\mhalf{b}-\mhalf{d},\mhalf{n}-m)(1+\mhalf{b}-\mhalf{a}-\mhalf{d},\mhalf{n}-m)(\mhalfo+\mhalf{a}+\mhalf{b}-\mhalf{d},m-\mhalf{n})}{(\mhalf{3},m)(\mhalfo-\mhalfo{a}+\mhalf{b}-\mhalf{d},-\mhalf{n}-m)(1-\mhalf{b}+\mhalf{b}-\mhalf{d},-\mhalf{n}-m)(c-\mhalf{a}-\mhalf{b}+\mhalf{d}-\mhalfo,\mhalf{n}-m)}\Bigg\rbrace.
\end{align}
This expression could already be used in calculations. Nonetheless it proves convenient to rewrite the sums for easier implementation in a framework for treating hypergeometric series. Therefore the half-integer values have to be transformed into integer ones. This can be achieved by making the replacements $\mhalf{n} \rightarrow n$ and $n \rightarrow n+\mhalfo$ for the even and the odd parts respectively. Thereby one ends up with four different terms in which a new series called $K$ can be identified:
\begin{equation}\label{eq:def-K}
K(a,b,c,d;e,f,g,h;x,y):=\sum_{m,n=0}^\infty \frac{(a,m+n)(b,m+n)(c,m-n)(d,m-n)}{(e,m-n)(f,m-n)(g,m)(h,n)}\frac{x^m}{m!}\frac{y^n}{n!}.
\end{equation}
The result for the analytically continued series, called $L$, is
\begin{align}
\label{eq:L}
&L(a,b;c,d;\frac{(1-y)^2}{4x},\frac{x}{4}):=\nonumber\\
	&\quad=\frac{\Gamma(d)\Gamma(a+b-d)}
		{\Gamma(a)\Gamma(b)}\Gamma(c)\Gamma(\frac{1}{2})(-4x)^{\frac{d}{2}-\frac{a}{2}-\frac{b}{2}}\times\nonumber\\
	&\quad\times\Biggl\lbrace  \frac{1}
		{\Gamma(\frac{a}{2}+\frac{b}{2}-\frac{d}{2}+\frac{1}{2})\Gamma(c-\frac{a}{2}-\frac{b}{2}+\frac{d}{2})}\times \nonumber\\
	&\qquad \times K\left(\begin{matrix}
		\frac{b}{2}-\frac{a}{2}+\frac{d}{2},\frac{a}{2}-\frac{b}{2}+\frac{d}{2},\frac{a}{2}+\frac{b}{2}-\frac{d}{2},\frac{a}{2}+\frac{b}{2}-c-\frac{d}{2}+1;\\
		\frac{b}{2}-\frac{a}{2}+\frac{d}{2},\frac{a}{2}-\frac{b}{2}+\frac{d}{2},\frac{1}{2},\frac{1}{2};
		\end{matrix}
		\frac{(1-y)^2}{4x},\frac{x}{4}\right)+\nonumber\\
	&\qquad+\frac{(d+a-b-1)(d-a+b-1)(-x)^{\frac{1}{2}}}{2(d-a-b+1)\Gamma(\frac{a}{2}+\frac{b}{2}-\frac{d}{2})\Gamma(c-\frac{a}{2}-\frac{b}{2}+\frac{d}{2}+\frac{1}{2})}\times\nonumber\\\
	&\qquad K\left(\begin{matrix}
		\frac{b}{2}-\frac{a}{2}+\frac{d}{2}+\frac{1}{2},\frac{a}{2}-\frac{b}{2}+\frac{d}{2}+\frac{1}{2},\frac{a}{2}+\frac{b}{2}-\frac{d}{2}-\frac{1}{2},\frac{a}{2}+\frac{b}{2}-c-\frac{d}{2}+\frac{1}{2};\\
		\frac{b}{2}-\frac{a}{2}+\frac{d}{2}-\frac{1}{2},\frac{a}{2}-\frac{b}{2}+\frac{d}{2}-\frac{1}{2},\frac{1}{2},\frac{3}{2};
		\end{matrix}
		\frac{(1-y)^2}{4x},\frac{x}{4}\right)
	\Biggr\rbrace +\nonumber\\
	&\quad+\frac{\Gamma(d)\Gamma(a+b-d)}{\Gamma(a)\Gamma(b)}\Gamma(c)\Gamma(-\frac{1}{2})(1-y)(-4x)^{\frac{d}{2}-\frac{a}{2}-\frac{b}{2}-\frac{1}{2}}\times\nonumber\\
	&\times\Biggl\lbrace \frac{1}{\Gamma(\frac{a}{2}+\frac{b}{2}-\frac{d}{2})\Gamma(c-\frac{a}{2}-\frac{b}{2}+\frac{d}{2}-\frac{1}{2})} \times \nonumber\\
	&\qquad \times K\left(\begin{matrix}
		\frac{b}{2}-\frac{a}{2}+\frac{d}{2}+\frac{1}{2},\frac{a}{2}-\frac{b}{2}+\frac{d}{2}+\frac{1}{2},\frac{a}{2}+\frac{b}{2}-\frac{d}{2}+\frac{1}{2},\frac{a}{2}+\frac{b}{2}-c-\frac{d}{2}+\frac{3}{2};\\
		\frac{b}{2}-\frac{a}{2}+\frac{d}{2}+\frac{1}{2},\frac{a}{2}-\frac{b}{2}+\frac{d}{2}+\frac{1}{2},\frac{3}{2},\frac{1}{2};
		\end{matrix}
		\frac{(1-y)^2}{4x},\frac{x}{4}\right)+\nonumber\\
	&\qquad+\frac{(d+a-b)(d-a+b)(-x)^{\frac{1}{2}}}{2(d-a-b+1)\Gamma(\frac{a}{2}+\frac{b}{2}-\frac{d}{2}-\frac{1}{2})\Gamma(c-\frac{a}{2}-\frac{b}{2}+\frac{d}{2})} \times \nonumber\\
	&\qquad \times K\left(\begin{matrix}
		\frac{b}{2}-\frac{a}{2}+\frac{d}{2}+1,\frac{a}{2}-\frac{b}{2}+\frac{d}{2}+1,\frac{a}{2}+\frac{b}{2}-\frac{d}{2},\frac{a}{2}+\frac{b}{2}-c-\frac{d}{2}+1;\\
		\frac{b}{2}-\frac{a}{2}+\frac{d}{2},\frac{a}{2}-\frac{b}{2}+\frac{d}{2},\frac{3}{2},\frac{3}{2};
		\end{matrix}
		\frac{(1-y)^2}{4x},\frac{x}{4}\right)
	\Biggr\rbrace.
\end{align}
From \eref{eq:H-4F3} to \eref{eq:L} Eqs. (\ref{eq:Poch-multi})
to (\ref{eq:def-PochhammerWithGamma}) were used extensively. Eq. (\ref{eq:L}) is quite a lengthy expression and therefore Exton \cite{Exton:1994de} abbreviated it with 
$L(a,b,c,d;(1-y)^2/(4x),x/4)$ \cite{Exton:1994de}. In contrast to the above defined $G$ and $K$ series, the combinations of $x$ and $y$ are fixed in the definition of $L$. 
This is caused by the appearance of $x$ and $y$ in the prefactors to the $K$ series. $L$ differs from the result given in \cite{Exton:1994de}:
\begin{itemize}
\item The second and fourth term have an additional factor $\frac{1}{2}$.
\item The third argument of the second K series has an additional $-\frac{1}{2}$.
\item The second $(-4x)$ has an additional $-\frac{1}{2}$ in the exponent.
\end{itemize}
The final result for the Appell function is then
\begin{align}\label{eq:final-continued-F4}
F_4(a,b;c,d;x,y)=&\frac{\Gamma(d)\Gamma(d-a-b)}{\Gamma(d-a)\Gamma(d-b)} G(a,b,1-d+a,1-d+b;1-d+a+b,c;x,1-y)+\nnnl
	&+L\left(a,b,c,d;\frac{(1-y)^2}{4x},\frac{x}{4}\right).
\end{align}
\end{widetext}

\subsection{Regions of Convergence}
\begin{figure}
\epsfig{file=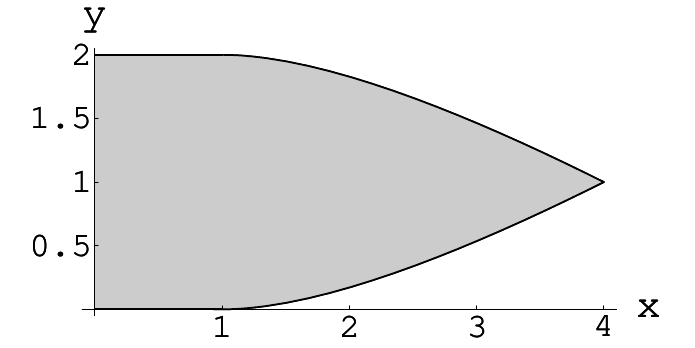,width=0.47\linewidth}
\epsfig{file=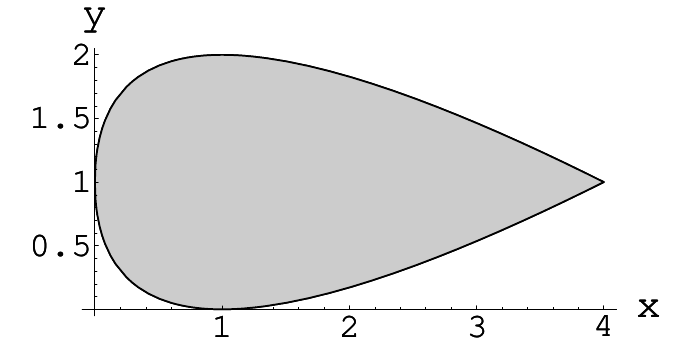,width=0.47\linewidth}
\caption{The left and right pictures shows the regions of convergence for the $G$ and $K$ series respectively for the variables as appearing in \eref{eq:final-continued-F4}. The region of convergence for the $K$ series is the smaller one and thus $K$ is the limiting series on the region of convergence.}
\label{fig:rocGK}
\end{figure}

Now we determine the regions of convergence. For $K$ we can employ the so-called method of cancellation of parameters. It can be shown \cite{Srivastava:1985hf} that the region of convergence does not depend on the parameters as long as these are not exceptional in the sense that the series is undefined, terminates or reduces to a sum of hypergeometric series of lower dimension. This means one can choose the parameters such that some Pochhammer symbols cancel each other. Choosing $c=e$ and $d=f$ the $K$ series reduces to the standard $F_4$ series for which the region of convergence is given in \eref{eq:F4-conv}. Inserting the occurring values for the variables the region of convergence for $K$, depicted in \fref{fig:rocGK} (right), is
\equ{\label{eq:roc-F4-continued}
\sqrt{\frac{(1-y)^2}{4x}}+\sqrt{\frac{x}{4}}<1.
}
For the $G$ series Horn's theorem on the convergence of double hypergeometric series \cite{Horn:1889ch} has to be used. For a short overview see refs. \cite{Erdelyi:1953ht,Srivastava:1985hf}. The resulting region of convergence is given in \fref{fig:rocGK} (left). One can see that it contains the region of convergence of the $K$ series and so the total region of convergence for \eref{eq:final-continued-F4} is given by \eref{eq:roc-F4-continued}.
There exist further analytic continuations with which one can cover the whole Euclidean momentum region. They are given in the appendix of \cite{Exton:1994de} and correspond to the analogous analytic continuations of the two additional series representations mentioned above (light gray regions in \fref{fig:roc}). In principle they are not necessary here, because we can always choose the ratios of the momenta such that the area in the rectangle defined by $(0,0)$ and $(1,1)$ is sufficient. Nonetheless we used several series in the numerical calculations because the convergence can be quite slow at the boundaries of the regions of convergence. One series used not given in \cite{Exton:1994de} is the one obtained by exchanging $c$ and $d$ simultaneously with $x$ and $y$. This is allowed because of the initial symmetry of $F_4$ under this exchange. One should note that $F_4$ is \textit{not} symmetric under the exchange of only $c$ and $d$ or $x$ and $y$. The resulting region of convergence is \eref{eq:roc-F4-continued} with $x$ and $y$ exchanged.

\subsection{Implementation of the Series}
\paragraph*{General case:}
In the form of \eref{eq:L} the series representation is not suitable for a numerical evaluation,
because it is possible that terms such as $\frac{0}{(0,m-n)}$ occur. Depending on the values of $m$ and $n$ this can be $0$ or a finite value. In order not to have to treat
all the possible cases of peculiar combinations on their own a slightly different form for $L$ was used. This is possible because some parameters of the $K$ series coincide. For the first and the third $K$ series these are the first/second and the fifth/sixth parameters. Exploiting this $K$ can be rewritten as
\begin{widetext}
\begin{align}\label{eq:def-K13}
&K_{1,3}(a,b,c,d;a,b,g,h;x,y)=\nnnl
	&\quad=\sum_{m,n=0}^\infty \frac{\G{a+m+n}\G{b+m+n}(c,m-n)(d,m-n)}{\G{a+m-n}\G{b+m-n}(g,m)(h,n)}\frac{x^m}{m!}\frac{y^n}{n!}.
\end{align}
In the second and fourth $K$ series the fifth/sixth parameter differs from the first/second by one. As the parameters also appear in the prefactor we can combine all this to
\begin{align}\label{eq:def-K24}
&2(a-1)\,2(b-1)\,K_{2,4}(a,b,c,d;a-1,b-1,g,h;x,y)=\nnnl
	&\quad=4\sum_{m,n=0}^\infty \frac{\G{a+m+n}\G{b+m+n}(c,m-n)(d,m-n)}{\G{a-1+m-n}\G{b-1+m-n}(g,m)(h,n)}\frac{x^m}{m!}\frac{y^n}{n!}.
\end{align}
\end{widetext}
Using Eqs. (\ref{eq:def-K13}) and (\ref{eq:def-K24}) we did not encounter the problems appearing for the original $K$ series.

\paragraph*{Special case:}
Normally Appell's function $F_4(a,b;c,d;x,y)$ is unity if one of the two first parameters vanishes. However, if it is multiplied by $\Gamma(b)$, and $b$ vanishes, the $1/\Gamma(b)$ in Appell's function is cancelled and the formerly vanishing terms become finite, whereas the first summand is $\Gamma(0)$. When calculating the ghost-gluon vertex exactly this happens for four integrals: The finite contributions become important and the $\Gamma(0)$ terms cancel. To get the finite parts one has to repeat parts of the derivation from above. The bottom line is that some single summands are no longer part of the series and the analytic continuation has to be done very carefully as it only works for complete series.

Right at the beginning we have instead of \eref{eq:F4-with-2F1}
\begin{widetext}
\begin{align}
\Gamma(b) & F_{4}(a,b;c,d;x,y)|_{b=0}= \Gamma(0)+\sum_{n=1}^{\infty}\frac{(a,n)}{(d,n)}\frac{y^{n}}{n}+\sum_{m=1}^{\infty}\frac{(a,m)\Gamma(m)}{(c,m)}\frac{x^{m}}{m!}\,_{2}F_{1}(a+m,m;d;y)
\end{align}
which leads to
\begin{align}
\Gamma(b) & F_{4}(a,b;c,d;x,y)|_{b=0}=\Gamma(0)+\frac{a\, y}{d}\,_{3}F_{2}(1,1,1+a;2,1+d;y)+M_0(a,0;c,d;x,y)
\end{align}
with
\begin{align}
M_0 & (a,0;c,d;x,y)= G_0(a,0,1-d+a,1-d;1-d+a,c;x,1-y)+L_0(a,b;c,d;\frac{(1-y)^{2}}{4x},\frac{x}{4})-\nnnl
& -\frac{\Gamma(d)\Gamma(a-d)}{\Gamma(a)}(1-y)^{d-a}\,_{2}F_{1}(d,d-a;1-a+d;1-y).
\end{align}
This result is not valid for $y=1$, in which case we have to use
\begin{align}
\Gamma(b) & F_{4}(a,b;c,d;x,1)|_{b=0}=\Gamma(0)+(\psi(d)-\psi(d-a))+M_0(a,0;c,d;x,1)
\end{align}
with
\begin{align}
M_0 & (a,0;c,d;x,1)= G_0(a,0,1-d+a,1-d;1-d+a,c;x,0)+L_0(a,0;c,d;0,\frac{x}{4})
\end{align}
and $\psi(x)$ the digamma function defined as $\Gamma'(x)/\Gamma(x)$. Additional auxiliary expressions are
\begin{align}
G_0(a,0,c,d;e,f;x,y)=\sum_{m=1,n=0}^{\infty}\frac{x^{m}}{m!}\frac{y^{n}}{n!}\frac{(a,m+n)\Gamma(m+n)(c,m)(d,m)}{(e,2m+n)(f,m)}
\end{align}
and
\begin{align}
L_0(a,b;c,d;\frac{(1-y)^{2}}{4x},\frac{x}{4})=\Gamma(b)L(a,b;c,d;\frac{(1-y)^{2}}{4x},\frac{x}{4}).
\end{align}
\end{widetext}

The structure of the regions of convergence is in this case more complicated and a more sophisticated algorithm to determine which one to choose is necessary.

\section{Tensor Bases for the Three-Gluon Vertex}\label{sec:tensorBases}

The most general basis for the three-gluon vertex consists of 14 tensors. They are build from all possible combinations of independent external momenta and the metric tensor. Under certain circumstances this number reduces. We list the used tensor bases here for reference.

The simplest case is when the three external momenta fulfill
\equ{\label{eq:def-symmPoint}
p_1^2=p_2^2=p_3^3=p^2,
}
which leads to
\equ{
p_1 \cdot p_2=p_2 \cdot p_3=p_3 \cdot p_1=-\frac{p^2}{2}.
}
This kinematic configuration is called the symmetric point and according to Celmaster and Gonsalves \cite{Celmaster:1979km} only three tensors are necessary:
\begin{align}\label{eq:tensors-symm}
&\Gamma^{(s)}_{\mu \nu \rho}(p_1,p_2,p_3)=\nnnl
	&=H_1(p^2)\left((p_1-p_2)_\rho \delta_{\mu\nu}+(p_2-p_3)_\mu \delta_{\nu\rho}+(p_3-p_1)_\nu \delta_{\mu\rho}\right)-\nnnl
	&\quad-H_2(p^2)\frac{(p_2-p_3)_\mu(p_3-p_1)_\nu(p_1-p_2)_\rho}{p^2}+\nnnl
	&\quad+H_3(p^2)\frac{p_{1_\rho}p_{2_\mu}p_{3_\nu}-p_{1_\nu}p_{2_\rho}p_{3_\mu}}{p^2}.
\end{align}
To get the scalars we contract the integral with the basis tensors and get the matrix equation
\begin{align}
&\tau^{(s)j}_{\mu \nu \rho}(p_1,p_2,p_3)\,\Gamma^{(s)}_{\mu \nu \rho}=\nnnl
&\sum_{i=1}^{3} H_i(p^2)\, \tau^{(s)j}_{\mu \nu \rho}(p_1,p_2,p_3)\,\tau^{(s)i}_{\mu \nu \rho}(p_1,p_2,p_3),
\end{align}
which can be solved for $H_1$, $H_2$ and $H_3$.
There are two ways to get numerical values for the contracted integrals: Either one decomposes all occurring scalars into $p^2$, $(q+p_1)^2$, $(q-p_2)^2$ and $q^2$ and calculates the integrals, or one uses the method described below to calculate the complete vertex and contracts the result with the basis tensors. For the symmetric point we cross-checked both methods.

The choice of the basis we used for the general kinematic case was motivated by two things: First it simplified calculations enormously, because it has its origin in the method described in \cite{Davydychev:1991va} to reduce tensor integrals to scalar ones and needs only one or two integrals for each tensor instead of several. Second, because some of the 14 original tensors coincide or the integrals are the same for different tensors, the number of necessary basis tensors decreases to 10:
\begin{align}
\Gamma_{\mu \nu \rho}^{\Delta}(p_1,p_2,p_3)=\sum_{i=1}^{10} E_i(p_1,p_2,p_3;\ka;d) \tau^i_{\mu \nu \rho}(p_1,p_2,p_3).
\end{align}
These are
\begin{subequations}
\label{eq:D10Tensors}
\begin{align}
\tau^1_{\mu \nu \rho}(p_1,p_2)&=\frac{p_{1_{\mu}}p_{1_{\nu}}p_{1_{\rho}}}{p_1^2},\\
\tau^2_{\mu \nu \rho}(p_1,p_2)&=\frac{p_{2_{\mu}}p_{2_{\nu}}p_{2_{\rho}}}{p_2^2},\\
\tau^3_{\mu \nu \rho}(p_1,p_2)&=\frac{p_{1_{\mu}}p_{1_{\nu}}p_{2_{\rho}}+p_{1_{\mu}}p_{2_{\nu}}p_{1_{\rho}}+p_{2_{\mu}}p_{1_{\nu}}p_{1_{\rho}}}{p_1^2},\\
\tau^4_{\mu \nu \rho}(p_1,p_2)&=\frac{p_{1_{\mu}}p_{2_{\nu}}p_{2_{\rho}}+p_{2_{\mu}}p_{1_{\nu}}p_{2_{\rho}}+p_{2_{\mu}}p_{2_{\nu}}p_{1_{\rho}}}{p_2^2},\\
\tau^5_{\mu \nu \rho}(p_1,p_2)&=g_{\mu \nu} p_{1_{\rho}}+g_{\mu \rho} p_{1_{\nu}}+g_{\nu \rho} p_{1_{\mu}},\\
\tau^6_{\mu \nu \rho}(p_1,p_2)&=g_{\mu \nu} p_{2_{\rho}}+g_{\mu \rho} p_{2_{\nu}}+g_{\nu \rho} p_{2_{\mu}},\\
\tau^7_{\mu \nu \rho}(p_1,p_2)&=\frac{p_{1_{\mu}}p_{1_{\nu}}p_{2_{\rho}}}{p_1^2},\\
\tau^8_{\mu \nu \rho}(p_1,p_2)&=\frac{p_{1_{\mu}}p_{2_{\nu}}p_{2_{\rho}}}{p_2^2},\\
\tau^9_{\mu \nu \rho}(p_1,p_2)&=\frac{p_{1_{\mu}}p_{2_{\nu}}(p_2-p_1)_{\rho}+(p_2-p_1)_{\mu}p_{1_{\nu}}p_{2_{\rho}}}{p_1\cdot p_2},\\
\tau^{10}_{\mu \nu \rho}(p_1,p_2)&=g_{\nu \rho} p_{1_{\mu}}-g_{\mu \nu} p_{2_{\rho}}.
\end{align}
\end{subequations}
We partially normalized them so that all of them have the same mass dimension.

As this basis is the result of a mathematical method and not of physical considerations it does not have explicit symmetries in the momenta. However, as we are only interested in direct numerical results we stick to this basis and do not transform to another one.
The general formula for a tensor integral of rank $M$ in ref. \cite{Davydychev:1991va}, eq. (11), has to be slightly changed, because it was originally derived for Minkowski space and we use the Euclidean metric:
\begin{align}\label{eq:tensorsInt}
&I^{(N)}_{\mu_1 \ldots \mu_M}(k_1,\ldots,k_N; \nu_1, \ldots, \nu_N;d)=\nnnl
	&=\sum_{\substack{\lambda,\sigma_1,\ldots,\sigma_N\\2\lambda+\sum \sigma_i=M}}
	(-1)^M\left(2\right)^{2M-3\lambda} \pi^{M-\lambda} \times \nnnl
	& \times \big\lbrace[g]^\lambda [k_1]^{\sigma_1} \ldots [k_N]^{\sigma_N} \big\rbrace_{\mu_1 \ldots \mu_M}(\nu_1,\sigma_1) \ldots (\nu_N,\sigma_N) \times \nnnl
	&\times I^{(N)}(p_1,\ldots,p_N; \nu_1+\sigma_1, \ldots, \nu_N+\sigma_N;d+2(M-\lambda)).
\end{align}
This equation can be used for all tensor integrals of rank $M$ with $N$ legs as long as a formula for the scalar integral for arbitrary dimension $d$ and arbitrary exponents $\nu_i$ of the squared momenta appearing in the loop is at hand. For its definition in the case of the three-point integral see \eref{eq:scalar-3Point-Integral}.
For the use of \eref{eq:tensorsInt} a fixed momentum routing is necessary depicted in \fref{fig:gen-1loop}. To translate it to the momentum routing we have in \fref{fig:triangle} for the ghost triangle we set $k_3=0$, $k_2=-p_2$ and $k_1=p_1$. The ghost-ghost-gluon triangle has a different momentum routing and we set $k_3=0$ and replace $k_2$ by $p_{gh}$.
The term in brackets is the symmetric tensor combination of the metric tensor $g$ and the momenta $p_1$, $\ldots$, $p_N$. E.g. the tensor $\tau_3$ stems from $\lbrace p_1^2 \, p_2 \rbrace_{\mu\nu\rho}$. The value of $\lambda$ goes from $0$ to the integer part of $M/2$ and the value of the $\sigma_i$ from $0$ to $M$. Thereby the sum is restricted such that only combinations of the metric tensor and the momenta occur which allow to distribute all $M$ indices among them. The 16 terms we initially get for the ghost triangle can be combined to the ten tensors of \eref{eq:D10Tensors}. We amended the tensors by adding normalization factors to get the correct canonical dimension and facilitate the comparison with the results of \cite{Alkofer:2008jy}. The technical advantage of this basis is the low number of integrals, namely 14, necessary to be calculated. With the method described above for the symmetric configuration the number of integrals necessary at least quadruplicates (using the basis where every basis tensor consists of only one expression), because the contraction of a tensor with the integrals yields four terms. Changing the basis increases the number of integrals due to more complex basis tensors. E.g. the Ball-Chiu basis \cite{Ball:1980ax} needs $46\times 4=184$ integrals, of which a few may have the same values for the exponents and thus coincide. The number of integrals is not important in the most part of the Euclidean regime, but at the boundaries convergence of the sums becomes slow so that less integrals mean quite a speed-up in calculations.
\fig{ht}{fig:gen-1loop}{general-1loop,height=7cm}{Initial momentum routing when using the method of ref. \cite{Davydychev:1991va} for calculating tensor integrals. The $\nu_i$ denote the powers of the squared momenta.}

\bibliographystyle{utphys} 
\bibliography{literature}

\end{document}